\documentclass[%
 aip,
 amsmath,amssymb,
 reprint,%
]{revtex4-1}

\usepackage{dcolumn}
\usepackage{bm}
\usepackage[utf8]{inputenc}
\usepackage[T1]{fontenc}
\usepackage{mathptmx}
\usepackage{etoolbox}
\usepackage{amssymb}
\usepackage[utf8]{inputenc}
\usepackage{textcomp}
\usepackage{subfigure}
\usepackage{parskip}
\usepackage{graphicx}
\usepackage{verbatim}
\usepackage[version=4]{mhchem}
\usepackage{siunitx}
\usepackage{longtable,tabularx}
\usepackage{float}
\usepackage{multirow}
\usepackage{supertabular}
\usepackage{booktabs}
\usepackage{indentfirst}
\usepackage{setspace}
\usepackage{mathdots}
\usepackage{adjustbox}
\usepackage{makecell}
\usepackage{comment}
\usepackage{amsmath, bm}
\usepackage{diagbox}
\usepackage{enumitem}
\usepackage[T1]{fontenc}
\usepackage{subfigure}
\usepackage{gensymb}
\usepackage{fix-cm}
\makeatletter
\def\@email#1#2{%
 \endgroup
 \patchcmd{\titleblock@produce}
  {\frontmatter@RRAPformat}
  {\frontmatter@RRAPformat{\produce@RRAP{*#1\href{mailto:#2}{#2}}}\frontmatter@RRAPformat}
  {}{}
}%
\makeatother
\begin{document}
\preprint{AIP/123-QED}
\title{Hovering efficiency optimization of the cycloidal propeller with end plates}

\author{Han Zhen Li}
\affiliation{School of Aeronautics, Northwestern Polytechnical University, 127, You Yi Xi Lu, Xi'an, Shan Xi, P.R.China, 710072}

\author{Yu Hu}
\email{Julius\_hu@hotmail.com}
\altaffiliation{School of Aeronautics, Northwestern Polytechnical University, 127, You Yi Xi Lu, Xi'an, Shan Xi, P.R.China, 710072}
\thanks{*Corresponding author: Julius\_hu@hotmail.com}

\author{Lai Zhang}
\affiliation{School of Aeronautics, Northwestern Polytechnical University, 127, You Yi Xi Lu, Xi'an, Shan Xi, P.R.China, 710072}

\author{Hong Bo Sun}
\affiliation{School of Aeronautics, Northwestern Polytechnical University, 127, You Yi Xi Lu, Xi'an, Shan Xi, P.R.China, 710072}

\author{Xu Chao Zhang}
\affiliation{School of Aeronautics, Northwestern Polytechnical University, 127, You Yi Xi Lu, Xi'an, Shan Xi, P.R.China, 710072}



\date{\today}

\begin{abstract}
Cycloidal propellers are known for their omnidirectional vectored thrust, enabling smooth transitions between hovering and forward flight, making them ideal for unmanned aerial vehicles (UAVs) and electric vertical take-off and landing (eVTOL) aircraft. However, cycloidal propellers tend to have lower hovering efficiency compared to screw propellers. Adding end plates to the blade tips can enhance hovering efficiency by reducing blade tip vortices. But the impact of these end plates and the optimal design for cycloidal propellers incorporating them have not been thoroughly studied. This paper seeks to optimize hovering efficiency and develop design theories for cycloidal propellers with end plates. Extensive force measurement experiments are conducted to identify designs with optimal hovering efficiency. The sliding mesh technique is employed to solve the unsteady Reynolds-averaged Navier-Stokes (URANS) equations for a detailed analysis. Experimental results indicate that the designs with end plates generally achieve significantly better hovering efficiency than those without end plates. End plates help to maintain hovering efficiency, even though the blade aspect ratio is as small as 1.5. The designs with stationary end plates are superior to those with rotating end plates because rotation introduces additional torque caused by the friction force. Designs featuring thick end plates outperform those with thin end plates, as the rounded edges can eliminate end plate vortices. The best design features stationary thick end plates, a chord-to-radius ratio of 0.65, and a large pitching amplitude of 40 degrees. It achieves a hovering efficiency of 0.72 with a blade aspect ratio of 3, which is comparable to that of helicopters. In contrast, for the cases without end plates, the highest hovering efficiency is merely 0.54. 
\end{abstract}

\maketitle

%

\setlength{\LTleft}{0pt}   
\setlength{\LTright}{\fill} 
\section{\label{sec:level1}Nomenclature}
$A$ =        the disk area of the cycloidal propeller($m^2$)  \\
$AR$ =        aspect ratio of blade \\
$b$ =         blade span($m$)\\
$C$ =         blade chord length($m$)\\
${C_{Q}}$ =    torque coefficient,${C_{Q}} = \frac{Q}{\rho A{\omega}^2{R}^3}$ \\
${C_{T}}$ =    thrust coefficient ${C_{T}} = \frac{T}{\rho A{\omega}^2{R}^2}$ \\ 
$C/R$ =       blade chord to radius ratio  \\
$D$ =         diameter of the cycloidal propeller($m$)  \\
${DL}$ =      disk loading ($N/m^2$)  \\
$F_{h}$=      the horizontal force ($N$)  \\
$F_{v}$ =     the vertical force ($N$)  \\
$FM$ =        figure of merit. $FM=\frac{T}{P}\sqrt{\frac{T}{2{\rho}A}}=PL\sqrt{\frac{DL}{2\rho}}$  \\
$P$  =        power consumption ($W$)  \\
${PL}$ =      power loading ($N/W$)  \\
$Q$  =        total aerodynamic torque,$Q = Q_{b} + Q_{e}$ ($Nm$)  \\
$Q_{b}$ =     aerodynamic torque generated by blade ($Nm$)  \\
$Q_{e}$ =     aerodynamic torque generated by end plates ($Nm$)  \\
$R$   =       radius of a cycloidal propeller ($m$)  \\
$Re$   =      Reynolds number  \\
$RPM$  =      rotation speed ($r/min$)\\
$T$     =     the thrust of cycloidal propeller, $T=\sqrt{F_{h}^{2}+F_{v}^{2}}$ ($N$) \\
$t_{e}$ =     end plate thickness ($mm$) \\
$V_{y}$ =     downwash velocity ($m/s$)  \\
$\alpha_{p}$ =  blade pitching angle ($deg$)\\
$\alpha_{max}$ =  blade pitching amplitude ($deg$)\\
$\delta_t$ =      the clearance between the blade tip and end plate ($mm$)\\
$\omega$ =    the angular velocity of the cycloidal propeller ($rad/s$)  \\
$\theta$  =       azimuth angle of the propeller($^\circ$)  \\
$\rho$    =       air density ($kg/m^3$)  \\
$\varphi$  =         azimuth angle of the resultant force ($deg$) \\

\section{Introduction}
\begin{figure}
    \includegraphics[width=0.9\linewidth]{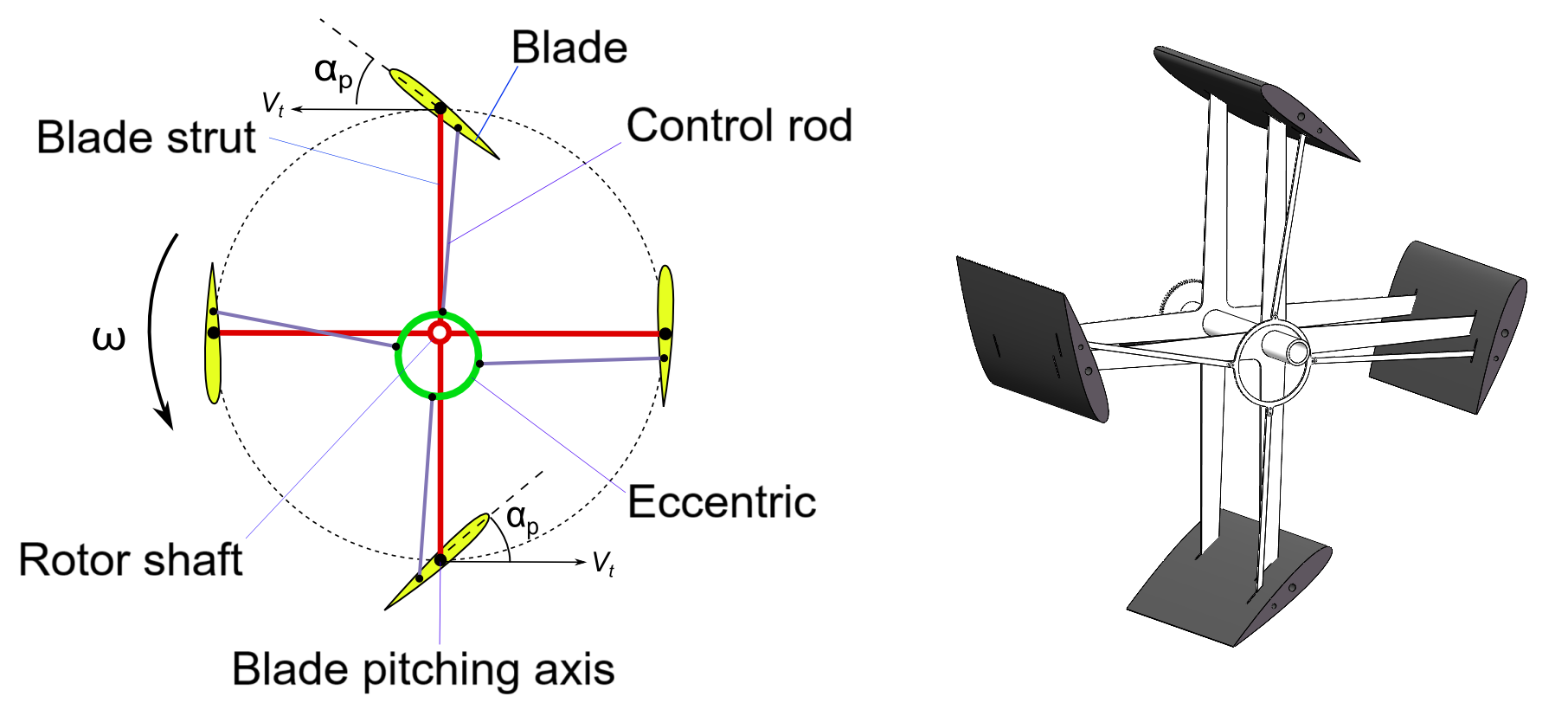}
    \caption{Illustration of a cycloidal propeller with four-bar control mechanism}
    \label{fig:Intro_Priciple_of_CYP}
\end{figure}

The cycloidal propeller is a type of propeller that can produce omni-directional thrust with low noise. It consists of several blades rotating around an axis that is parallel to the rotor shaft. The periodic pitch angle of each blade is controlled by a control rod and an eccentric, as shown in Figure.\ref{fig:Intro_Priciple_of_CYP}. If the position of the eccentric is changed, the magnitude and direction of the thrust will change instantly. This feature enables a seamless transition between hover and forward flight. This enables the aircraft to perform exceptionally well in tasks such as vertical takeoff and landing (VTOL), low-speed flight, and hovering\cite{Akttaktt:2024}. The fact that each blade element works at an identical velocity allows cycloidal propellers to produce thrust at a low rotation speed while also generating less noise. This leads to a quieter UAV and eVTOL aircraft . Despite these advantages, cycloidal propellers tend to have lower hovering efficiency compared to screw propellers. As a result, numerous efforts are underway to enhance the hovering efficiency of cycloidal propellers.

Based on the four-bar control mechanism and sinusoidal blade pitching motion, extensive research has been conducted to improve the aerodynamic efficiency of the cycloidal propeller \cite{Hu:2015a,Hu:2019a,Louis_2021,Zimmer_Louis:2023, HU2024118289,Tang2018}. Sun et al. developed a non-symmetric pitch-control model, and their research indicated that imposing different pitch amplitudes between the upper and lower half-cycles can enhance hovering efficiency \cite{ZZSun:2025a}. Hu and Wan performed extensive 2D numerical simulations on cycloidal propellers \cite{HU2024118289}. Their research shows that the longer the blade chord, the greater the thrust. Additionally, the blade thickness has little effect on side thrust; however, it can significantly increase the efficiency of the propeller. Previous studies revealed that the chord-to-radius ratio, represented as $C/R$, is the most important factor influencing hovering efficiency. A $C/R$ in the range of $0.6$ to $0.7$ results in the most efficient cycloidal propellers, with hovering efficiency as high as $0.78$ \cite{Hu:2015a}. A higher blade aspect ratio can improve efficiency, whereas decreasing the taper ratio typically reduces hovering efficiency, as it leads to a decreased blade chord-to-radius ratio \cite{Hu:2019a}. Zimmer and Gagnon studied how the Reynolds number impacts the aerodynamics of the cycloidal propeller \cite{Zimmer_Louis:2023}. Their findings indicated that regardless of the flow regime, rotation-averaged lift and power followed the predictions of momentum theory, except for $Re$ of 500,000. Saito and Kurose recently  investigated the effect of interactions between the blades and the blade-tip vortex on the aerodynamic performance of a cyclorotor using Large Eddy Simulation \cite{SAITO2024108921,SAITO2025109951}. The results show that the blade-tip vortex effect increases the thrust generated by the blades on the blowing side of the cyclorotor. Furthermore, when the number of blades exceeds a specific value, the thrust generated by each blade significantly decreases, and total thrust generation begins to decline.

Some scholars and research teams have also introduced a variety of unconventional control mechanisms and blade designs aimed at enhancing the performance of cycloidal propellers. Ferrier introduced cycloidal propellers with active leading-edge morphing \cite{Ferrier_Vezza_Zare-Behtash_2017,Ferrier:2019}. This study showed that applying active leading-edge morphing resulted in significant improvements in the cycloidal propeller’s performance characteristics. Fasse and Sacher et al. optimized the pitching oscillation of the blade by controlling the motion of the individual blade using servo motors \cite{FASSE2024117363}. Compared to traditional sinusoidal pitching motion, this optimized pitching motion results in a hydro efficiency gain of $10\%$ to $20\%$. M. Habibnia and J. Pascoa \cite{Mehdi:2021a,HABIBNIA2021107141} optimized the pitching motion of the blades based on Artificial Neural Networks (ANN). Their study demonstrated that by implementing real-time control of the pitch angle oscillation of the blades, a substantial enhancement in the hovering efficiency of the cycloidal propeller could be attained. Lei Shi \cite{SHI2024118393} proposed augmenting the performance of cycloidal propellers by deflecting the blade leading and trailing edges, thereby achieving an increase in both side force and propulsive force. Benmoussa and Páscoa,\cite{BENMOUSSA2021106468} investigated the integration of Dielectric Barrier Discharge (DBD) plasma actuators with cycloidal propeller blades. This research indicated that the employment of external actuation during the upper half-cycle and internal actuation during the lower half-cycle effectively enhanced the lift on the cycloidal propeller blades, increasing the peak lift value per blade in $11\%$ and the total thrust by $1.5\%$.

\begin{figure}
    \centering
    \subfigure[Potential $FM$ the cycloidal propeller may achieve]
    {
        \includegraphics[width=0.5\linewidth]{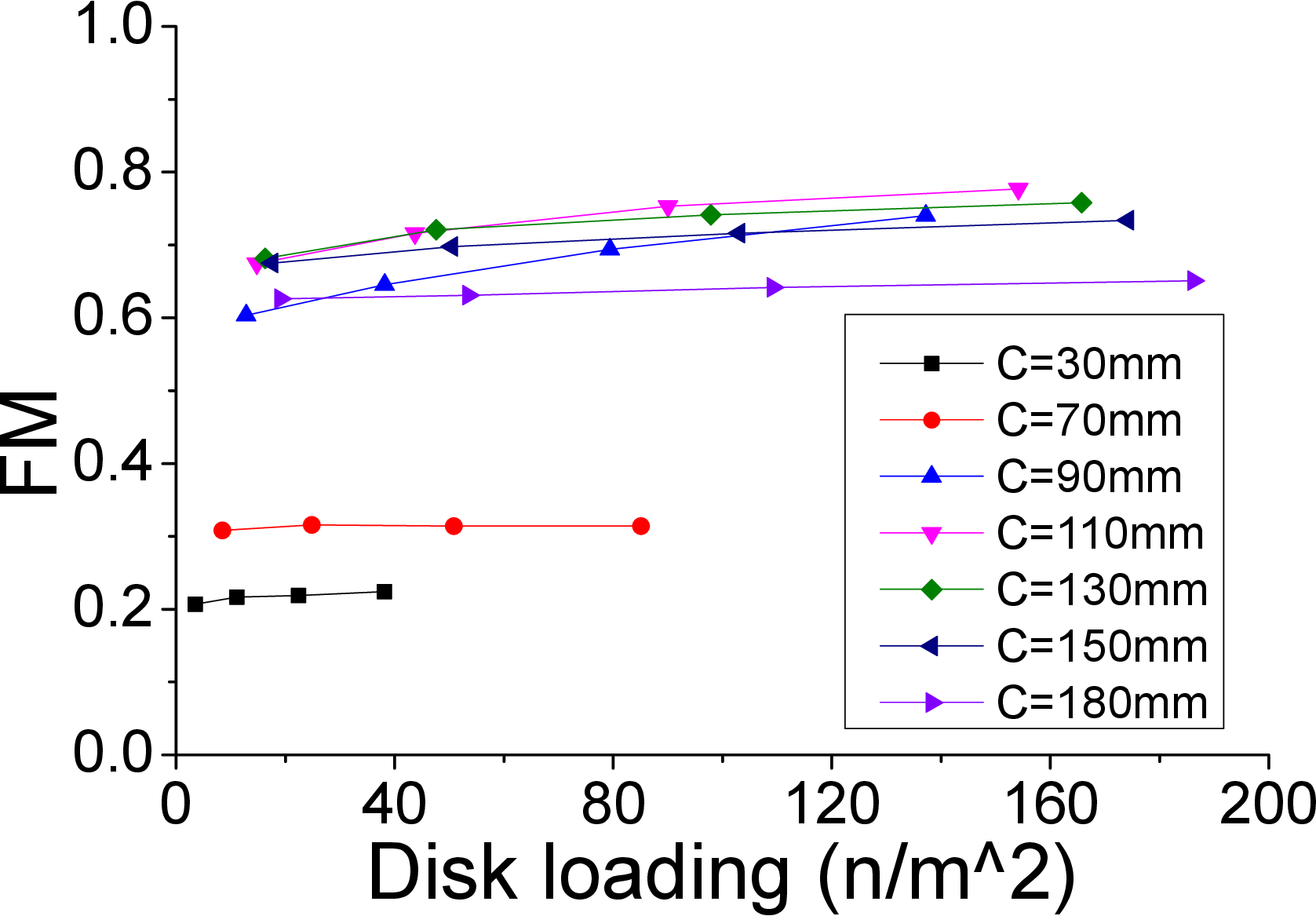}
        \label{fig:2D_CFD_CR}
    }
    \subfigure[The cycloidal propeller with end plates]
    {
        \includegraphics[width=0.4\linewidth]{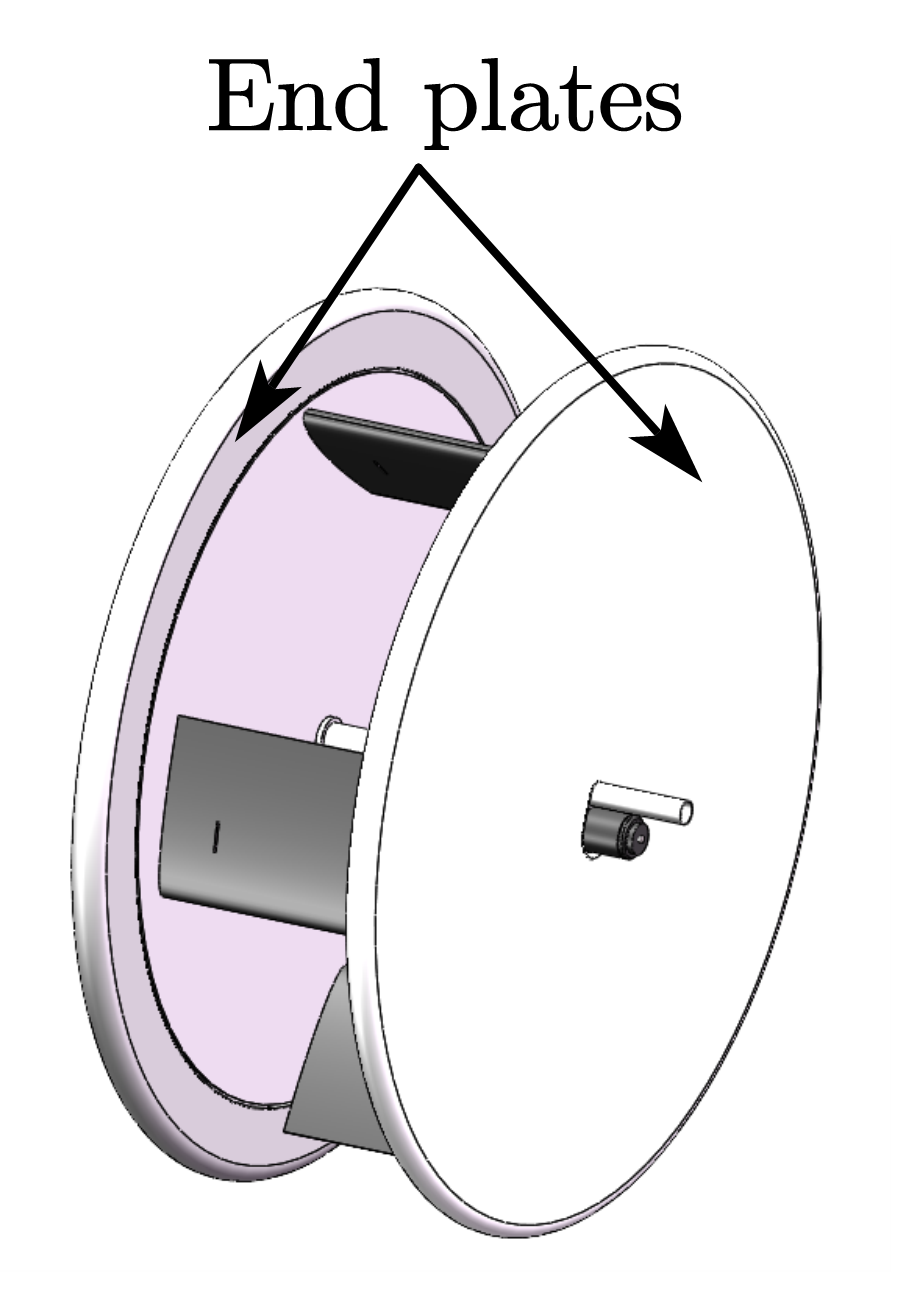}
        \label{fig:Cyp_With_End_plates}
    }
    \caption{2D numerical simulation results for cycloial propeller with different blade chord length \cite{Hu:2015a} and the cycloidal propeller with end plates}
    \label{fig:2D_CFD_CR_Cyp_With_End_plates}
\end{figure}

Existing two-dimensional numerical simulations indicate that the hovering efficiency can potentially reach up to $0.78$ if the blade chord to radius ratio is high\cite{Hu:2015a}, as illustrated in Figure \ref{fig:2D_CFD_CR}. However, these simulations do not account for the loss due to blade tip vortices, which can significantly reduce hovering efficiency. By incorporating two end plates at the blade tips, as shown in Figure \ref{fig:Cyp_With_End_plates}, we can expect an improvement in efficiency, as the end plates help to constrain the blade tip vortices. In this case, it is also possible to select a low blade aspect ratio to reduce structural weight. However, the aerodynamics of cycloidal propellers equipped with end plates have yet to be thoroughly investigated, and there are currently no established design principles for enhancing the hovering efficiency of these cycloidal propellers.

This paper aims to optimize the hovering efficiency of cycloidal propellers equipped with end plates and to formulate design theories for these configurations. Our study examines cycloidal propellers featuring end plates and low blade aspect ratios, specifically in the range of \(3 \geq AR \geq 0.5\). For the sake of simplicity, we employ a conventional sinusoidal pitching motion introduced by a four-bar mechanism.

The contributions of this paper are as follows: 

\begin{enumerate}[label=(\arabic*). ]
\item Cycloidal propellers with end plates and low blade aspect ratio are thoroughly studied based on force measurement experiments. Numerical simulations based on URANS are performed for detailed analysis. 
\par
\item The designs with and without end plates are studied and compared. 
\par 
\item The impact of blade chord to radius ratio, blade aspect ratio, the effects of end plates and end plate rotation, as well as end plate thickness are studied. 
\par
\item Based on experimental and numerical simulations, the most efficient designs have been identified. 

\end{enumerate}

\section{The experiment apparatus} 
\begin{figure}
    \centering
    \subfigure[The cycloidal propeller assembly]
        {
            \includegraphics[width=0.6\linewidth]
            {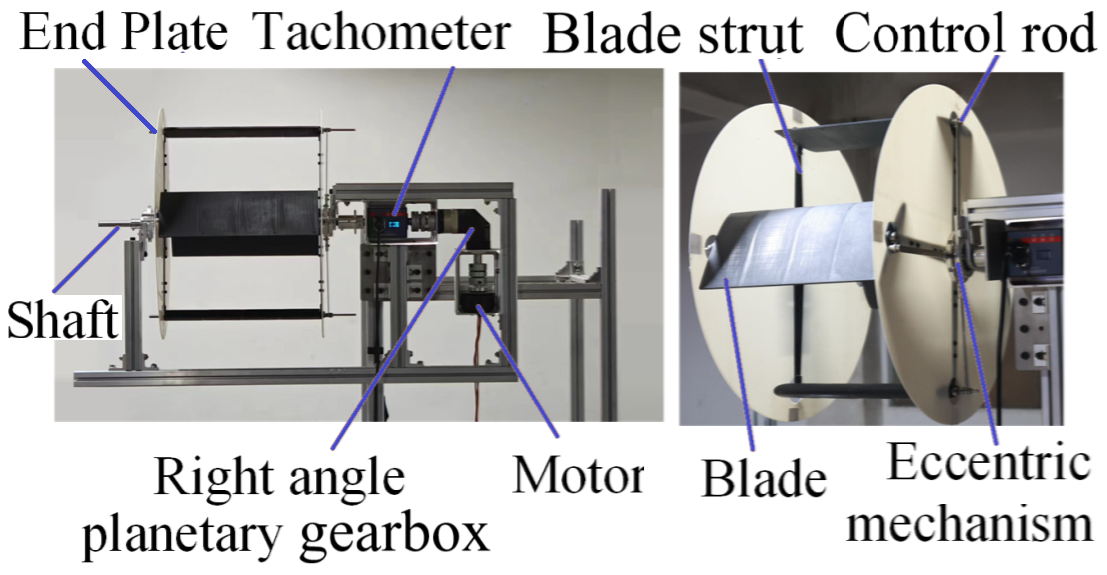}
            \label{fig:Experiment_Apparatus_2}
        }
    \\    
    \subfigure[Exploded view of the experiment apparatus]
        {
            \includegraphics[width=0.45\linewidth]
            {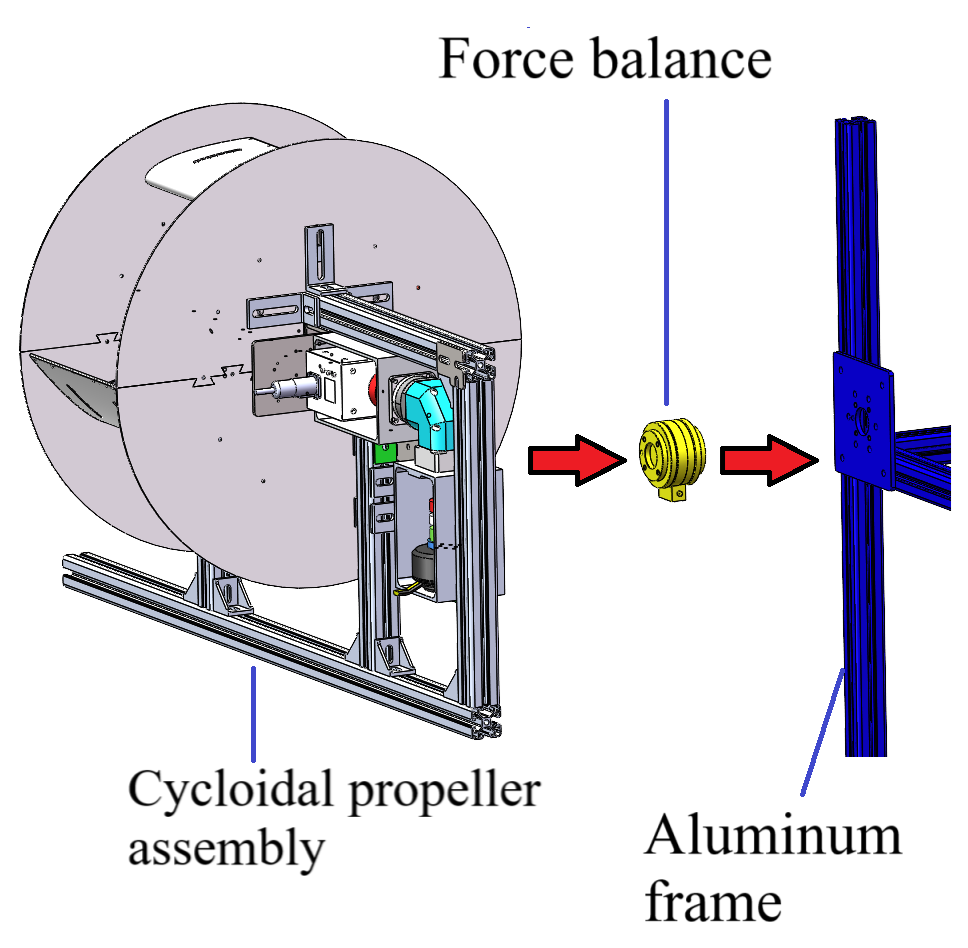}
            \label{fig:Experiment_Apparatus_3}
        }
    \subfigure[The cycloidal propeller with fixed and thick end plates]
        {
            \includegraphics[width=0.4\linewidth]
            {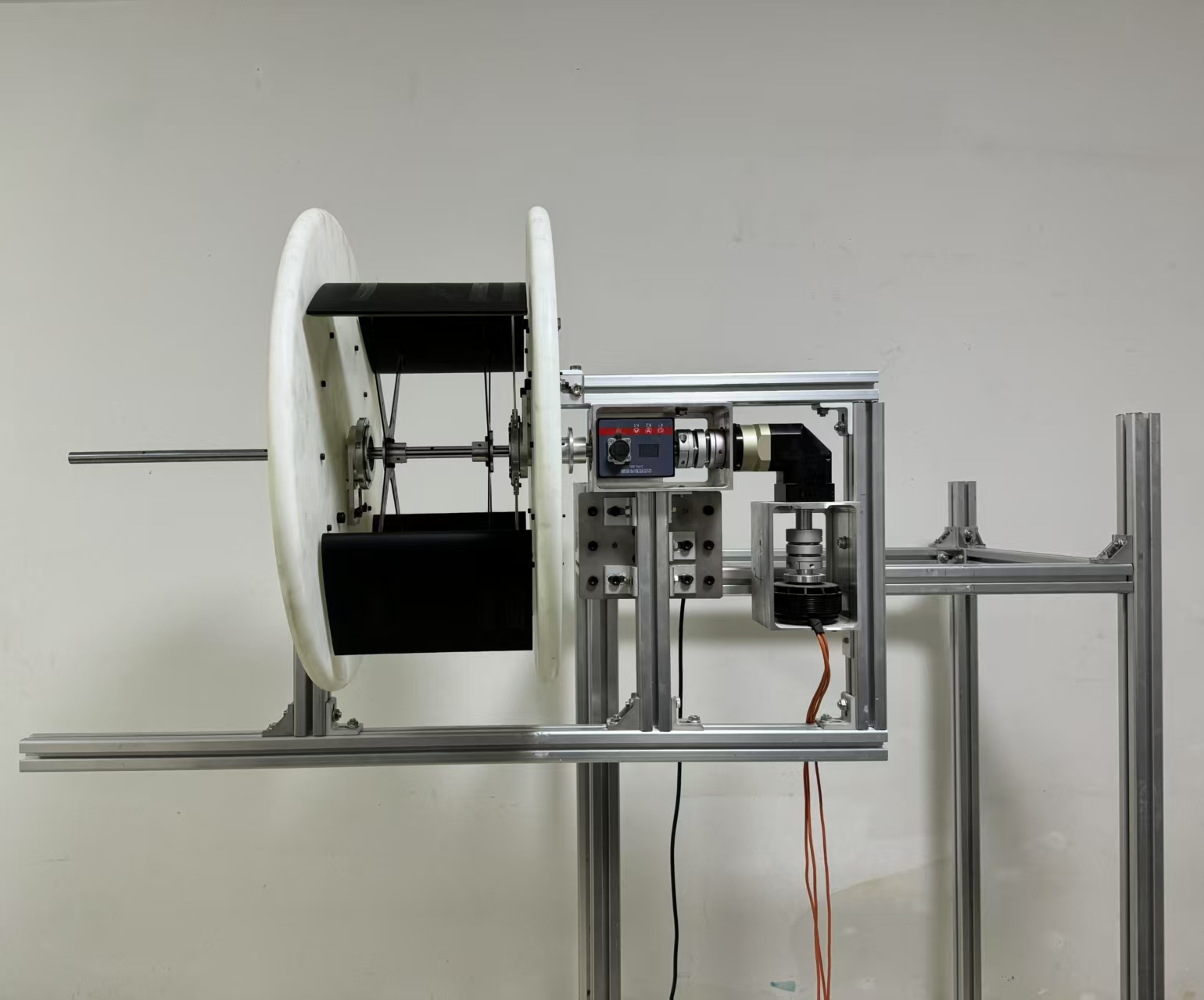}
            \label{fig:Experiment_Apparatus_4}
        }
    \caption{Experimental setup for cycloidal propellers}
    \label{fig:Experiment_Apparatus}
\end{figure}

Figure.\ref{fig:Experiment_Apparatus} illustrates the experimental setup, consisting of a cycloidal propeller assembly, a force balance, and an aluminum support frame. The entire system is supported by the aluminum frame. The force balance links the cycloidal propeller assembly to the aluminum frame, allowing for the measurement of the aerodynamic forces generated by the cycloidal propeller. The propeller thrust and slipstream are aligned horizontally to prevent interference with the ground. Experiments are conducted in the center of a laboratory hall. During the experiments, the cycloidal propeller is placed at least 10 meters away from the wall, which is over 20 times the diameter of the rotor, to minimize any interference from the walls.

The cycloidal propeller assembly consists of the cycloidal propeller itself and an aluminum rack. This rack stabilizes both ends of the cycloidal propeller, helping to reduce vibrations caused by rotor shaft deformation. Additionally, the aluminum rack also serves as the mounting structure for the electric motor, tachometer, stationary end plates, and reduction gearbox.

Two removable end plates are attached near the tip of the blade. Two sets of cycloidal propeller structures have been developed so that the end plates can either rotate with the rotor or be fixed onto the aluminum rack to remain stationary. It is powered by a brushless motor with a maximum power of $3kW$. A right angle planetary speed reducer with a gear ratio of 3:1 is used to constrain the maximum rotation speed within $2000RPM$. A tachometer is installed between the output shaft of the speed reducer and the cycloidal propeller shaft. The cycloidal propeller and the rack are fixed on the six-component force balance. The resolution of the X, Y force components is $0.3N$ and $0.88N$ respectively, and the resolution of the torques is $0.06Nm$.  

\section{The numerical simulation model}
All numerical simulations are conducted using ANSYS Fluent. Since the blades in all cases operate at velocities of less than $30 m/s$, which falls into the low-speed range, the incompressible URANS equations are solved. The pressure implicit with the splitting of operators (PISO) scheme is utilized to solve the pressure-velocity coupling problem. The convective and diffusive flux terms are discretized using a second-order upwind scheme and a second-order central difference scheme, respectively. For the transient formulation, a second-order implicit time-marching algorithm is applied. The order of the Reynolds number is approximately $10^5$, so the shear stress transport (SST) model with low Reynolds correction is employed. This model can predict boundary layer transition and dynamic stall of the airfoil with greater accuracy.

\begin{figure}
    \centering
    \subfigure[Definition of the coordinate system]
    {
        \includegraphics[width=0.5\linewidth]{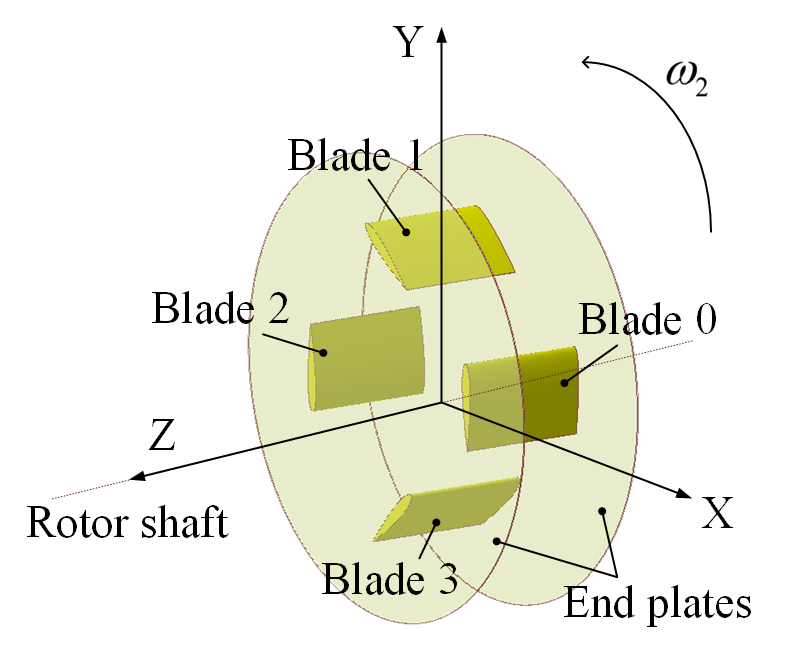}
    }
    \\
    \subfigure[Definition of the forces in inertial frame]
    {
        \includegraphics[width=0.45\linewidth]{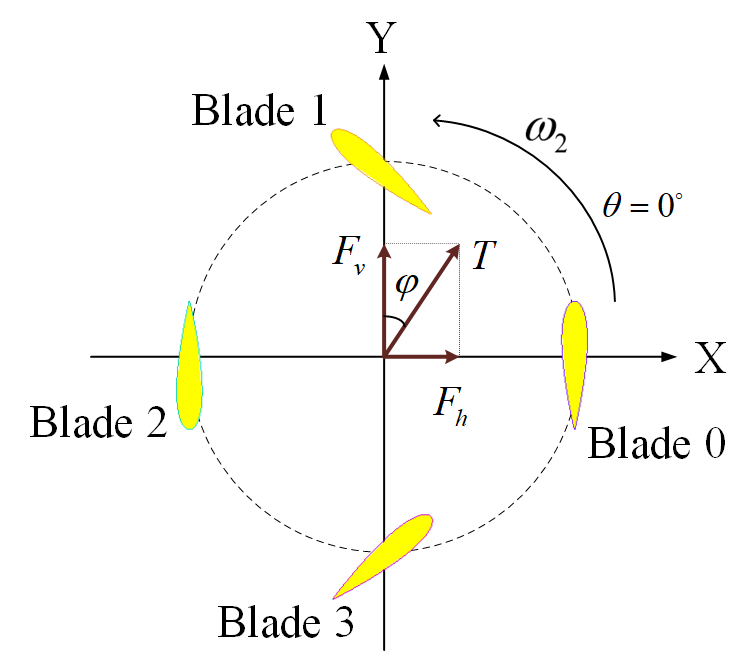}
    }
    \subfigure[Definition of the forces in local frame]
    {
        \includegraphics[width=0.45\linewidth]{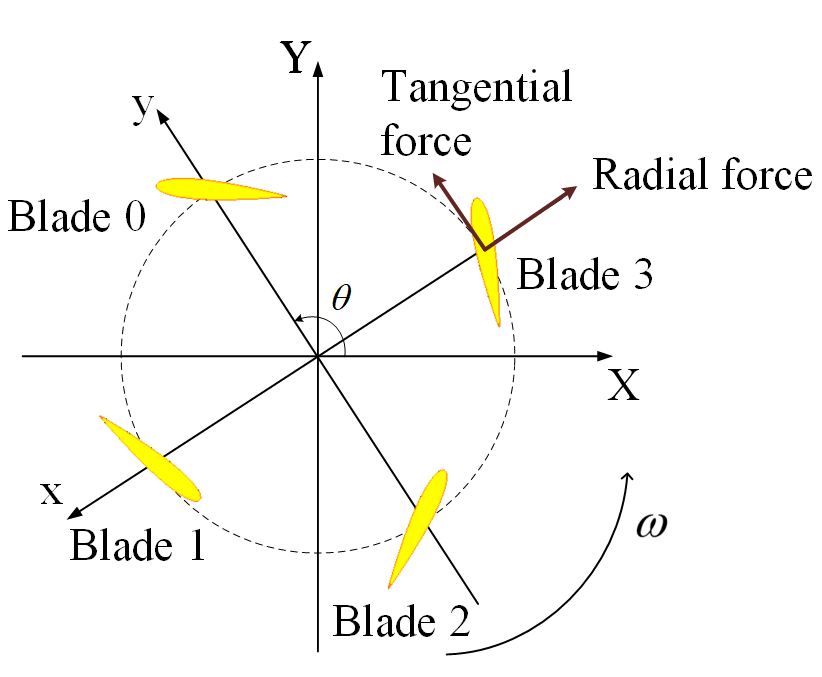}
    }
    
    \caption{Definitions of reference frames and aerodynamic forces}
    \label{fig:coordinate_system}
\end{figure}

The coordinate system employed in the numerical simulations is shown in Figure.\ref{fig:coordinate_system}. The cycloidal propeller rotates along the Z axis. The motion of the blade can be divided into two components: the rotation around the rotor shaft and the pitching oscillation about the blade's pitch axis. To model this motion, the sliding mesh technique is employed. This approach eliminates the need for mesh deformation or re-meshing at each time step, thereby avoiding low-quality elements that arise from mesh deformation and reducing the time required for mesh updates.

\begin{figure}
    \centering
    \subfigure[The organization of the multiple stage sliding mesh system]
    {
        \includegraphics[width=0.85\linewidth]{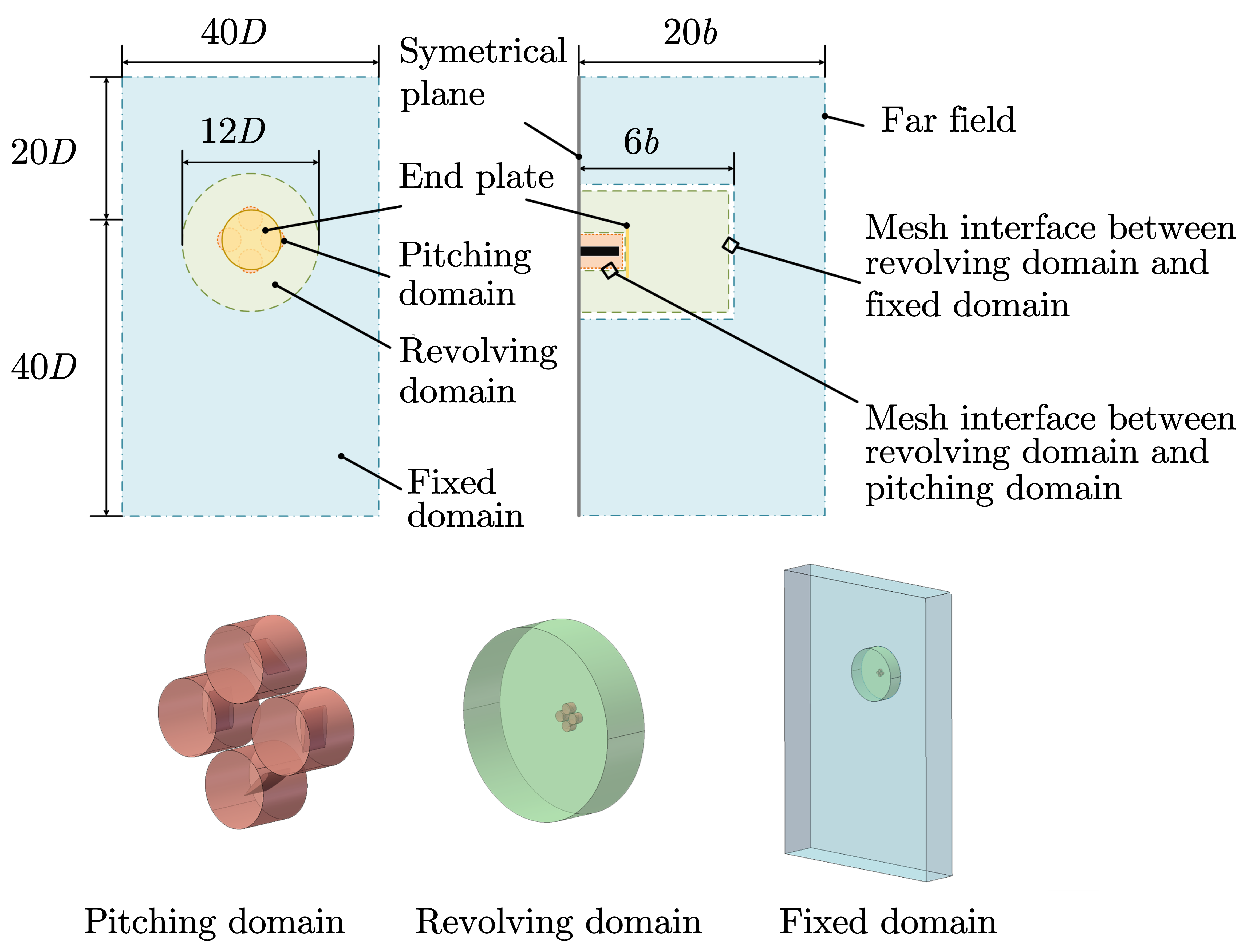}
    }
    \subfigure[Surface mesh of the cycloidal propeller with end plates]
    {
        \includegraphics[width=0.65\linewidth]{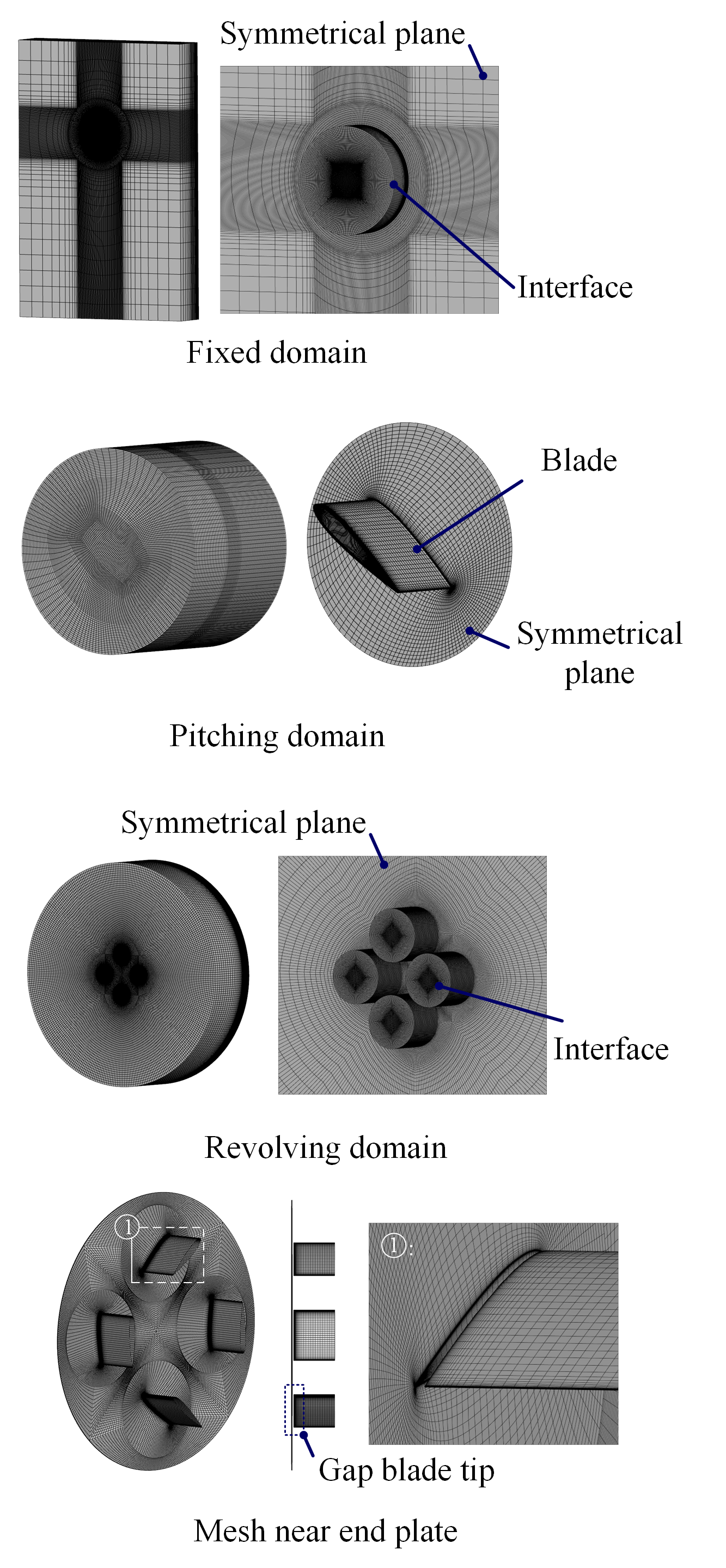}
    }
    \caption{The sliding mesh system for cycloidal propellers with end plates}
    \label{fig:Moving_mesh_skeme}
\end{figure}

As shown in Figure.\ref{fig:Moving_mesh_skeme}, three levels of mesh blocks are deployed in the computation domain. A cylindrical hole is cut in the fixed domain. The revolving domain around the cycloidal propeller is embedded in that hole. The cylindrical hole for each blade is then cut in the revolving domain. The pitching domain around each blade is fitted to the corresponding hole in the revolving domain. Mesh elements are condensed around the blade surface, the blade tip, and the regions where the blade wakes and vortices are expected. For all cases, the sizes of the two mesh elements adjacent to the mesh interfaces are carefully set so that the ratio of their sizes is less than 2.0. This can reduce the influence of the hanging nodes, which cause discontinuities and spurious vortex structures.

The structured mesh is deployed throughout the entire computation domain for all cases. Only the cycloidal propeller with a half-blade span is modeled, since the flow is symmetric about the mid-plane. In this way, the number of mesh elements is reduced by half. The far-field condition is set to zero pressure gradients and zero velocity. In order to simulate the boundary layer effects more accurately, no wall functions are used. Hence, the height of the mesh of the first row near the blade surface is set such that $y^{+}$ is equal to or less than 1. In directions perpendicular to the rotor shaft, the far field stretches at least 20 times the diameter of the rotor. In the direction parallel to the rotor shaft, the dimension of the computational domain is stretched to at least 20 times the blade span. This ensures that the influence of the far field is minimized.

The mesh independent experiments are carried out. The results obtained with different mesh sizes are validated against experimental data. It is found that for each pitching domain around the blade, 240 mesh elements in the chordwise direction and 54 mesh elements in the radial direction are enough to generate a mesh-independent solution. Together, there are about 4 million to 7 million mesh elements in the computational domain, depending on the $AR$. Various time step lengths are tested, and finally, it is found that 1500 steps per cycle, or a Courant number of around 3, are enough to produce a time-independent solution.

\section{Results and discussion}
\subsection{Validation of numerical simulation model}

\begin{figure}
    \centering
    \subfigure[Thrust of case C9N]
    {
        \includegraphics[width=0.47\linewidth]{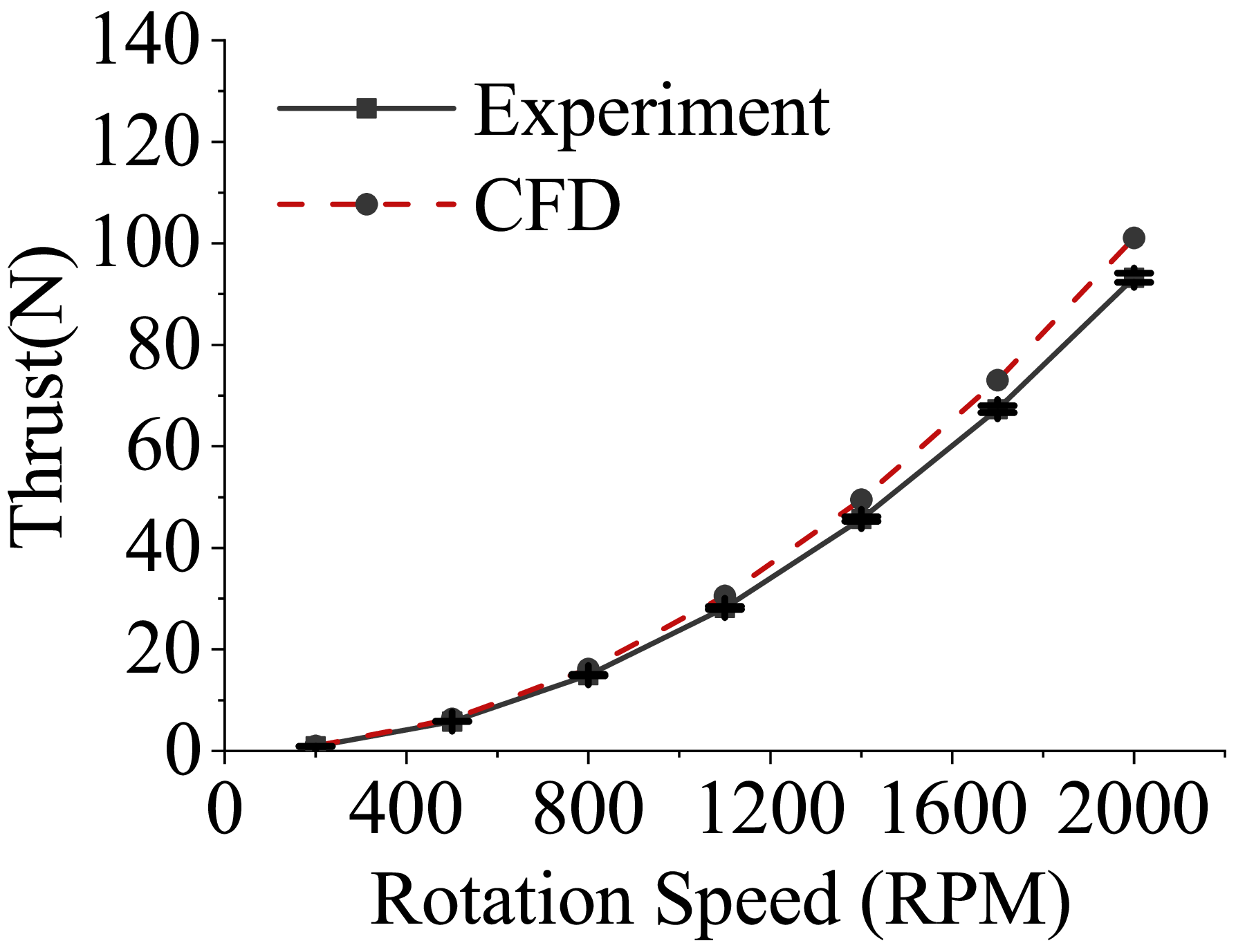}
    }
    \subfigure[Thrust of case C9D]
    {
        \includegraphics[width=0.47\linewidth]{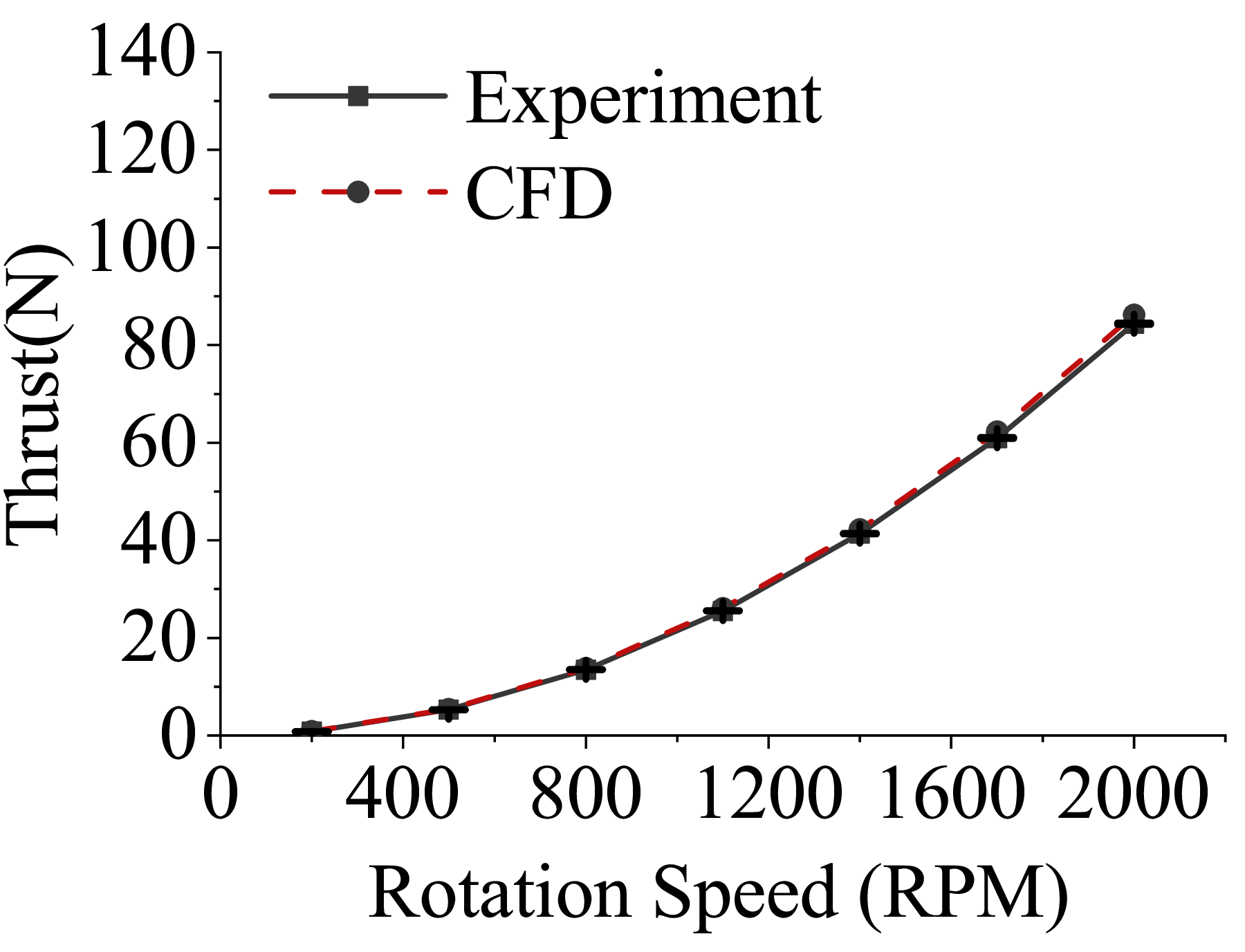}
    }
    \\
    \subfigure[Torque of case C9N]
    {
        \includegraphics[width=0.47\linewidth]{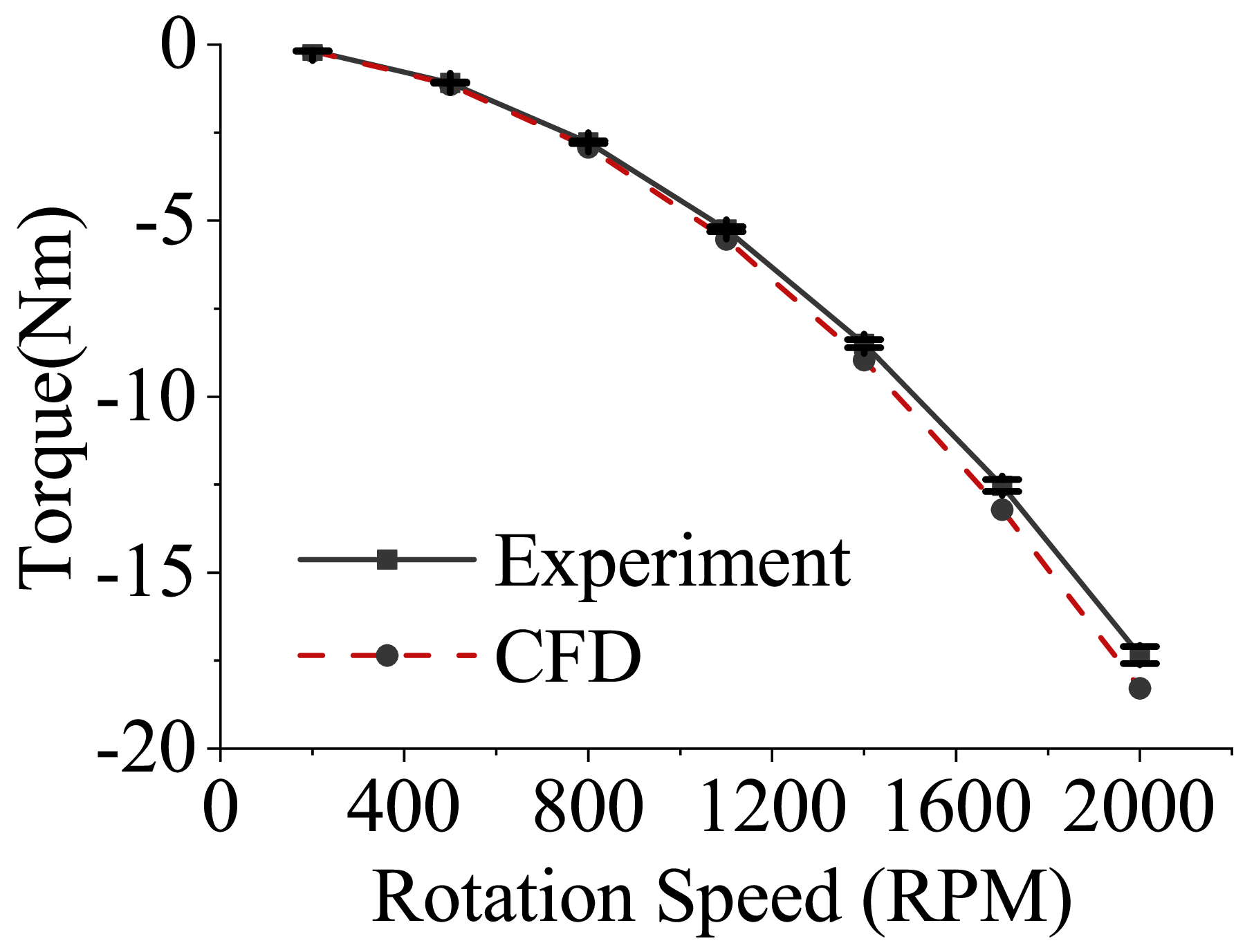}
    }
    \subfigure[Torque of case C9D]
    {
        \includegraphics[width=0.47\linewidth]{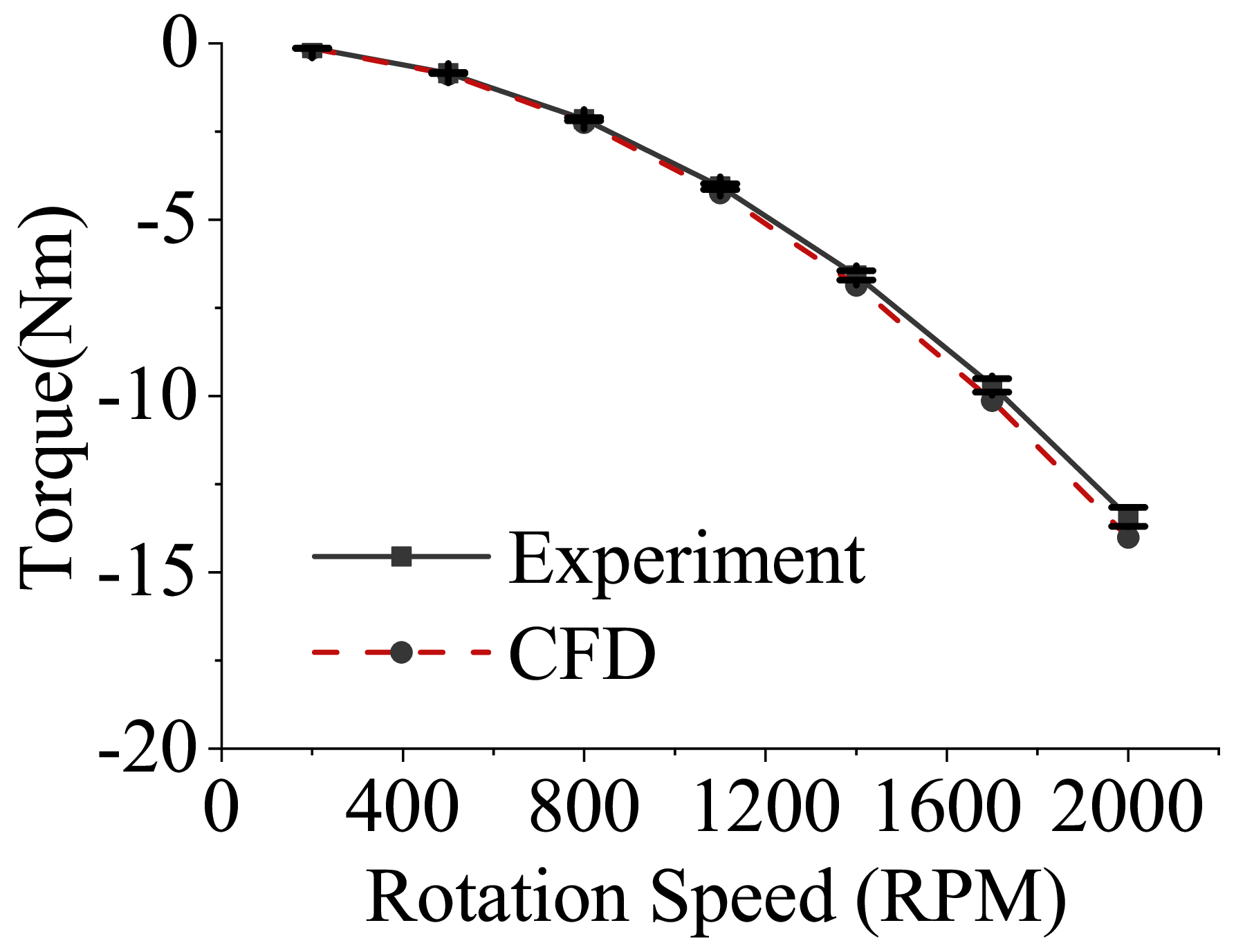}
    }
    \subfigure[$FM$ of case C9N]
    {
        \includegraphics[width=0.47\linewidth]{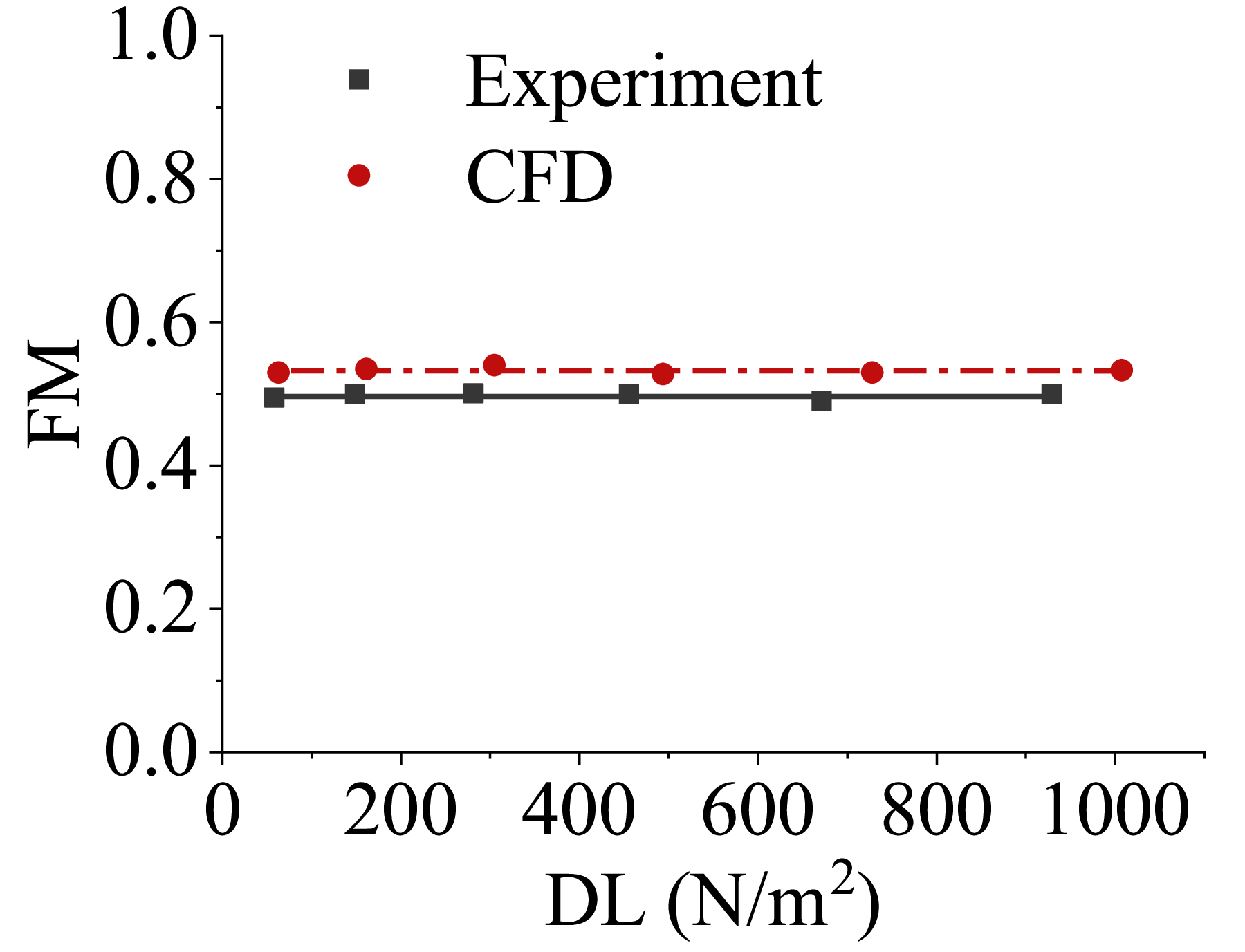}
        \label{fig:Case9N_CFDandEx}
    }
    \subfigure[$FM$ of case C9D]
    {
        \includegraphics[width=0.47\linewidth]{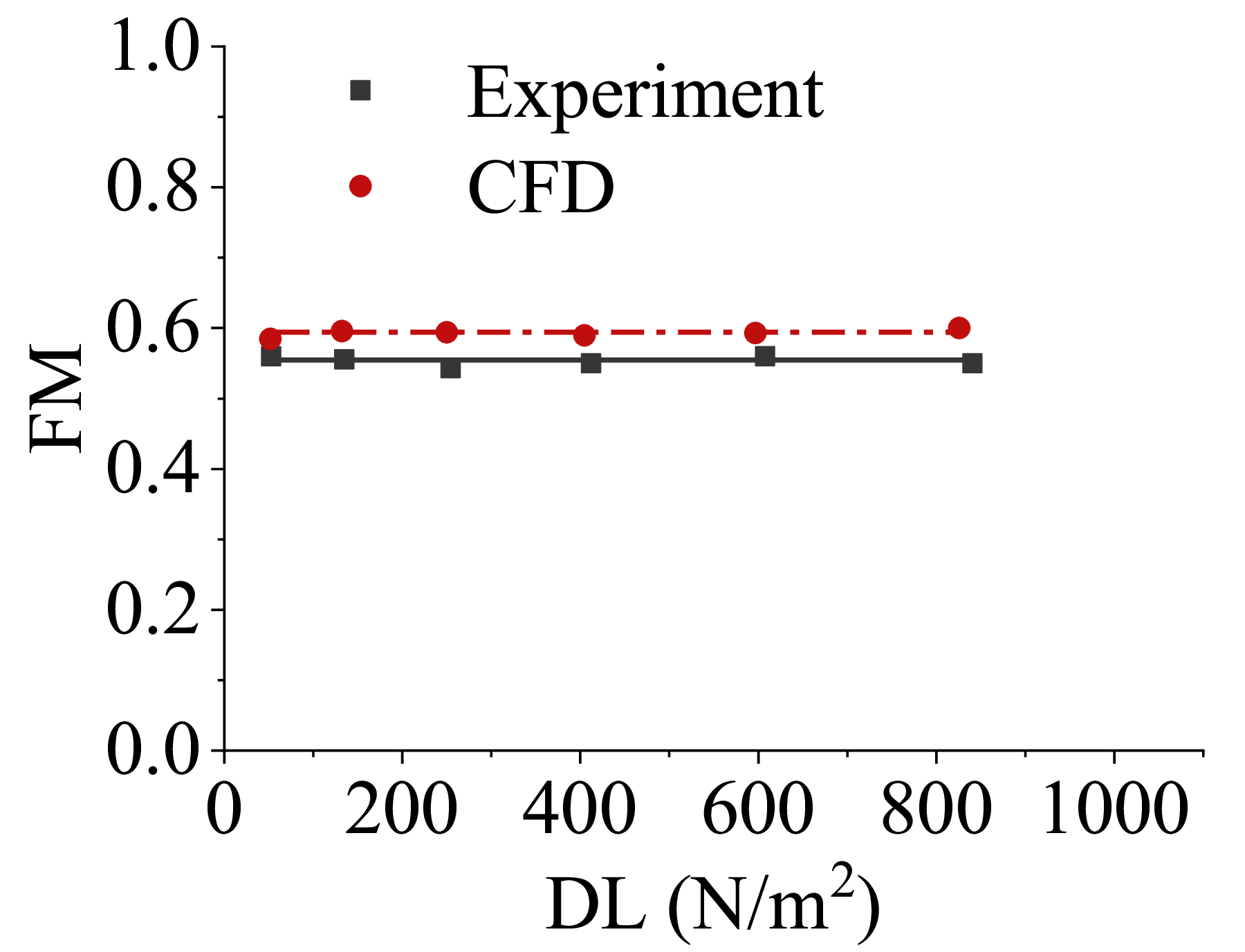}
        \label{fig:Case9D_CFDandEx}
    }
    \caption{Comparison of numerical simulation and experiment results for the cases defined in Table.\ref{table:Experiment_Cases_CR}}
    \label{fig:Exp_CFD_Comparison}
\end{figure}

The numerical simulation models proposed in this article are validated using experimental data as shown in Figure.\ref{fig:Exp_CFD_Comparison}. It can be seen that the model based on sliding mesh can produce quite good predictions for the cases with and without end plates. 

\subsection{Effects of blade chord to radius ratio for the cases with and without rotating end plates}

In this section, experiments are performed on the basis of the cycloidal propellers with and without end plates. 

The test cases with different $C/R$ are listed in Table \ref{table:Experiment_Cases_CR}. The rotor diameters of all test cases are $450mm$, and the blade numbers are 4. For each configuration, the blade pitching amplitudes from $10^{\circ}$ to $50^{\circ}$ are also tested. The IDs of these cases are followed by a suffix to state the number of end plates, where $N$ stands for no end plates and $D$ stands for double end plates. For example, "C1D" means that it is case 1 with double end plates, and "C1N" means case 1 without end plates. 

\begin{table}[ht]\centering
\caption{Parameters of the cycloidal propellers with different $C/R$}
\begin{tabular}{ccccc}
\hline
Case ID & $C/R$ & $AR$ \\
\hline
C1$\left[N,D\right]$   & 0.26      & 3.81\\
C2$\left[N,D\right]$   & 0.36      & 2.75\\
C3$\left[N,D\right]$   & 0.46      & 2.15\\
C4$\left[N,D\right]$   & 0.56      & 1.77\\
C9$\left[N,D\right]$   & 0.65      & 1.50\\
\hline
N: no end plates, D: double rotating end plates
\label{table:Experiment_Cases_CR}
\end{tabular}
\end{table}

\begin{figure}
    \centering
    \includegraphics[width=0.8\linewidth]{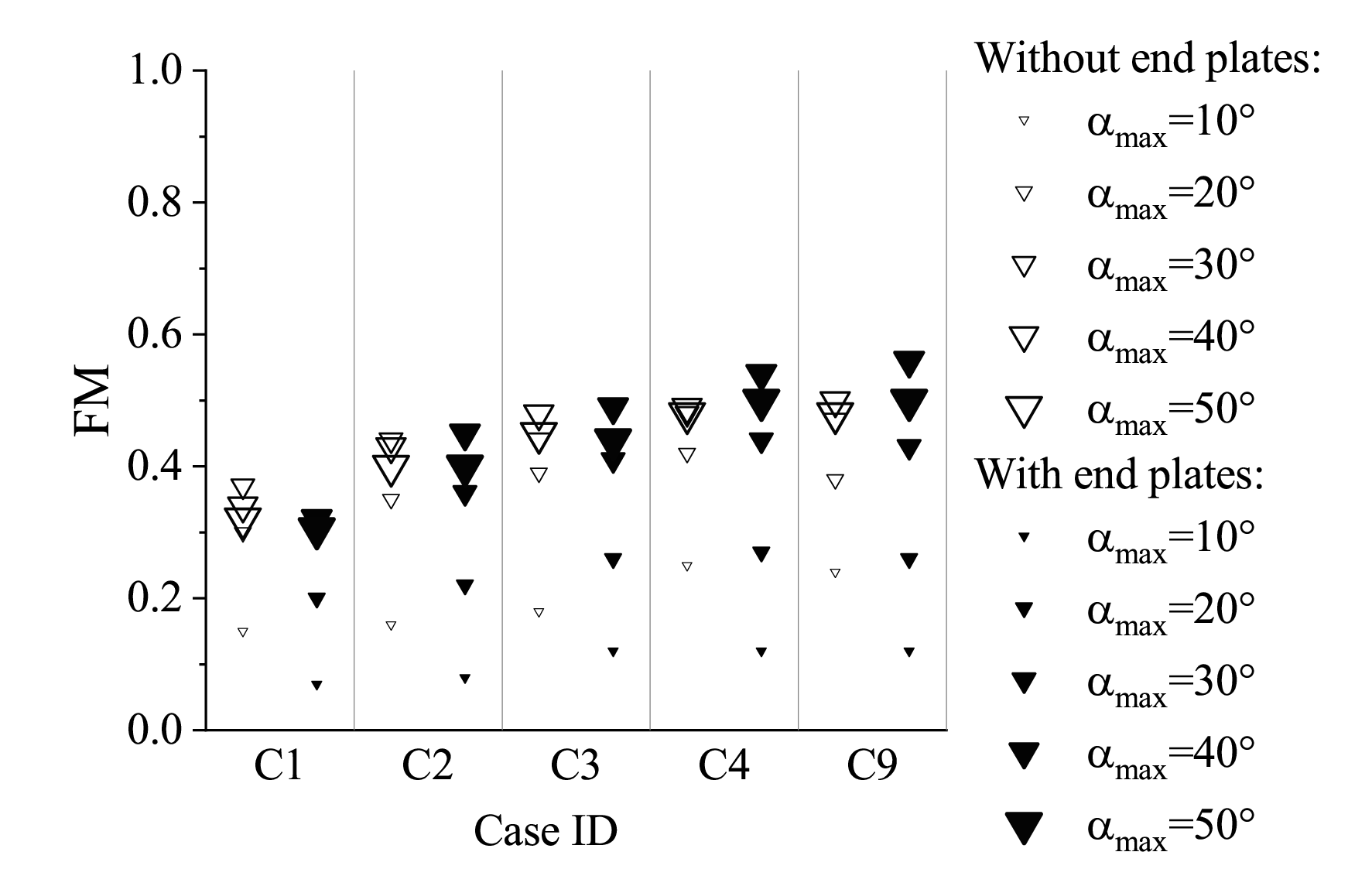}
    \caption{Effects of $C/R$ on hovering efficiency}
    \label{fig:Case_FM_All_CR}
\end{figure}

The hovering efficiencies of these cases are compared in Figure.\ref{fig:Case_FM_All_CR}. It can be seen that the cases with $C/R=0.65$ and pitching amplitude of $40^{\circ}$ prevail, regardless of whether there are end plates or not. The cases with rotating end plates perform much better than their counterparts when $\alpha_{max}=40^{\circ}$. 

\subsection{Effects of blade aspect ratio for the cases with and without rotating end plates}

\begin{figure}
    \centering
    \includegraphics[width=0.97\linewidth]{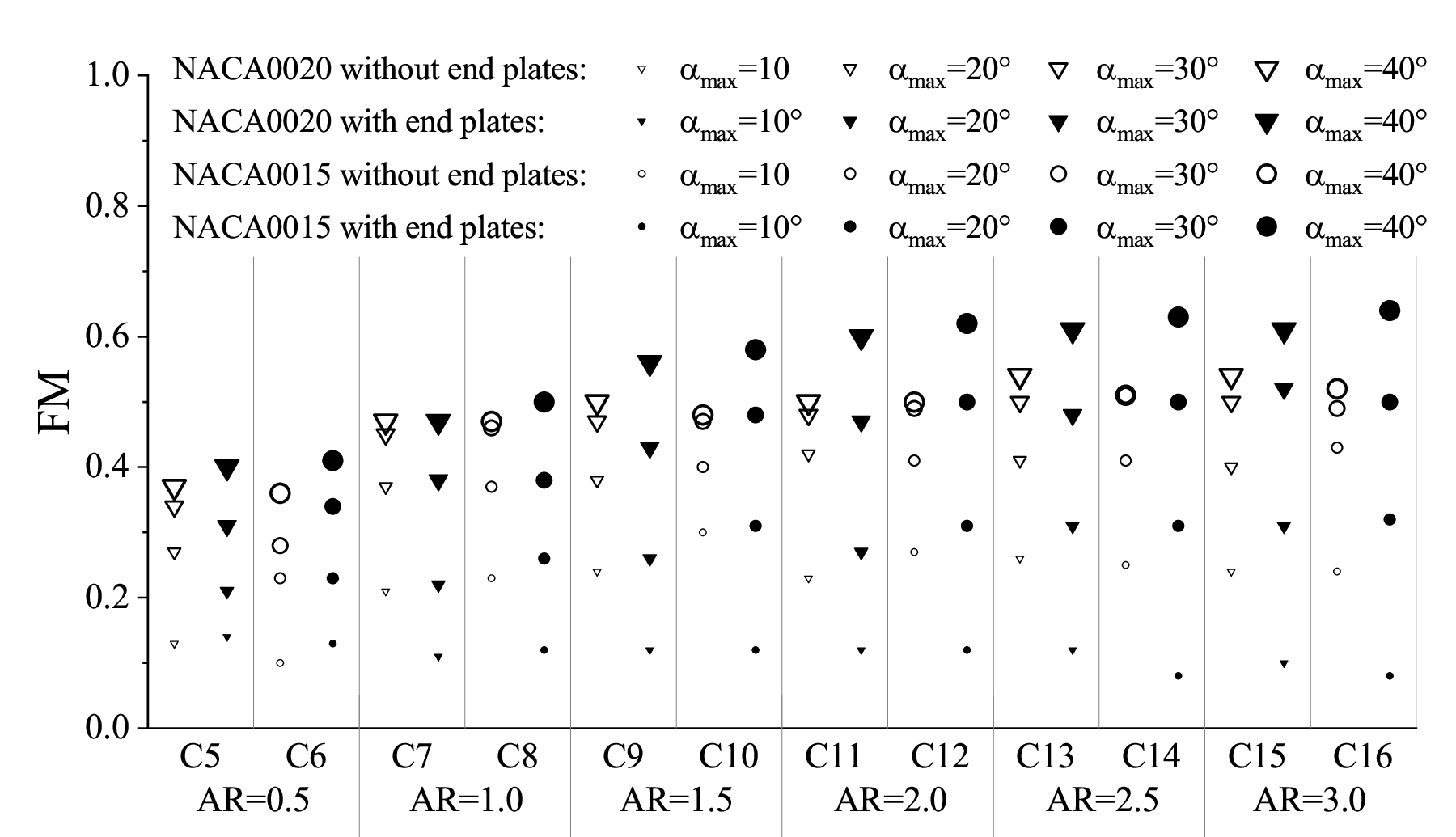}
    \caption{Effects of blade aspect ratio on hovering efficiency for cases with and without end plates}
    \label{fig:Effects_of_AR}
\end{figure}

The cases with different $AR$ are defined in Table.\ref{table:Experiment_Cases_AR}. The $C/R$ of these cases is $0.65$. The cases with and without rotating end plates are compared. For each configuration, the blade pitching amplitudes from $10^{\circ}$ to $40^{\circ}$ are also tested.  Their hovering efficiencies are compared in Figure.\ref{fig:Effects_of_AR}. 

\begin{table}[ht]\centering
\caption{Parameters of the cycloidal propellers with different $AR$}
\begin{tabular}{cccc}
\hline
Case ID & $AR$ & Airfoil \\
\hline
C5$\left[N,D\right]$  & 0.50    & NACA0020 \\
C6$\left[N,D\right]$  & 0.50    & NACA0015 \\
C7$\left[N,D\right]$  & 1.00    & NACA0020 \\
C8$\left[N,D\right]$  & 1.00    & NACA0015 \\
C9$\left[N,D\right]$  & 1.50    & NACA0020 \\
C10$\left[N,D\right]$ & 1.50    & NACA0015 \\
C11$\left[N,D\right]$ & 2.00    & NACA0020 \\
C12$\left[N,D\right]$ & 2.00    & NACA0015 \\
C13$\left[N,D\right]$ & 2.50    & NACA0020 \\
C14$\left[N,D\right]$ & 2.50    & NACA0015 \\
C15$\left[N,D\right]$ & 3.00    & NACA0020 \\
C16$\left[N,D\right]$ & 3.00    & NACA0015 \\
\hline
N: no end plates, D: double end plates
\label{table:Experiment_Cases_AR}
\end{tabular}
\end{table}

Figure.\ref{fig:Effects_of_AR} demonstrates that setups with double rotating end plates show superior performance compared to those without end plates. The optimal case is C16D, whose $FM=0.64$. In contrast, the highest $FM$ value for the cases without end plates is only $0.54$ (Case C15N). 

The designs with a pitching amplitude of $40^\circ$ yield the most favorable results. Moreover, it indicates that when double rotating end plates are present, the efficiency of hovering also decreases with a decrease in aspect ratio.

By comparing $FM$ of the cases with different airfoils, we can find that the NACA0020 airfoil outperforms NACA0015 when there are no end plates. However, for the cases with end plates, the NACA0015 airfoil is better. 

\begin{figure}
    \centering
    \subfigure[Torque produced by each component]
    {
        \includegraphics[width=0.45\linewidth]{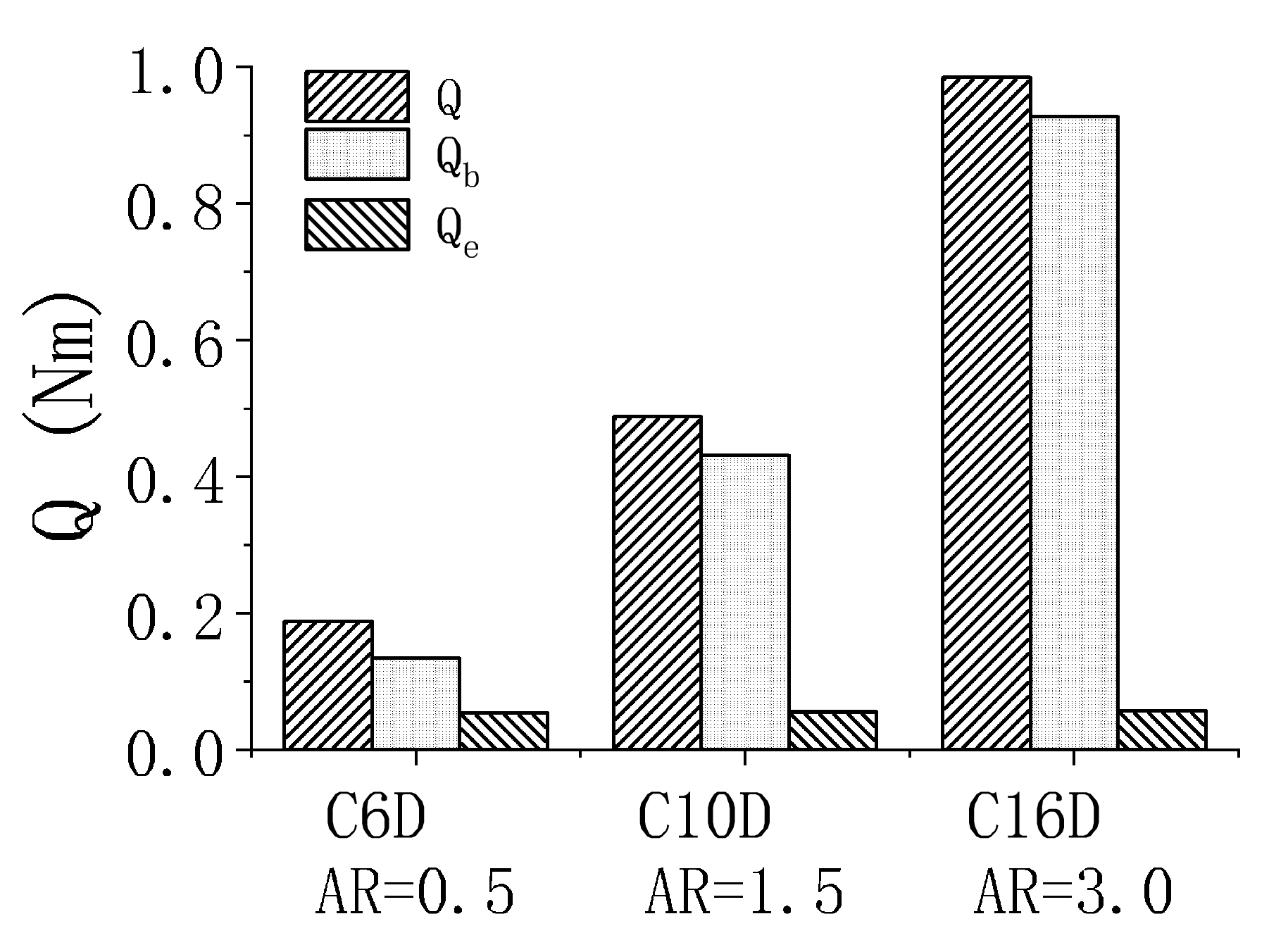}
        \label{fig:C6DC10DC16D_Torque}
    }
    \subfigure[Proportion of torques]
    {
        \includegraphics[width=0.45\linewidth]{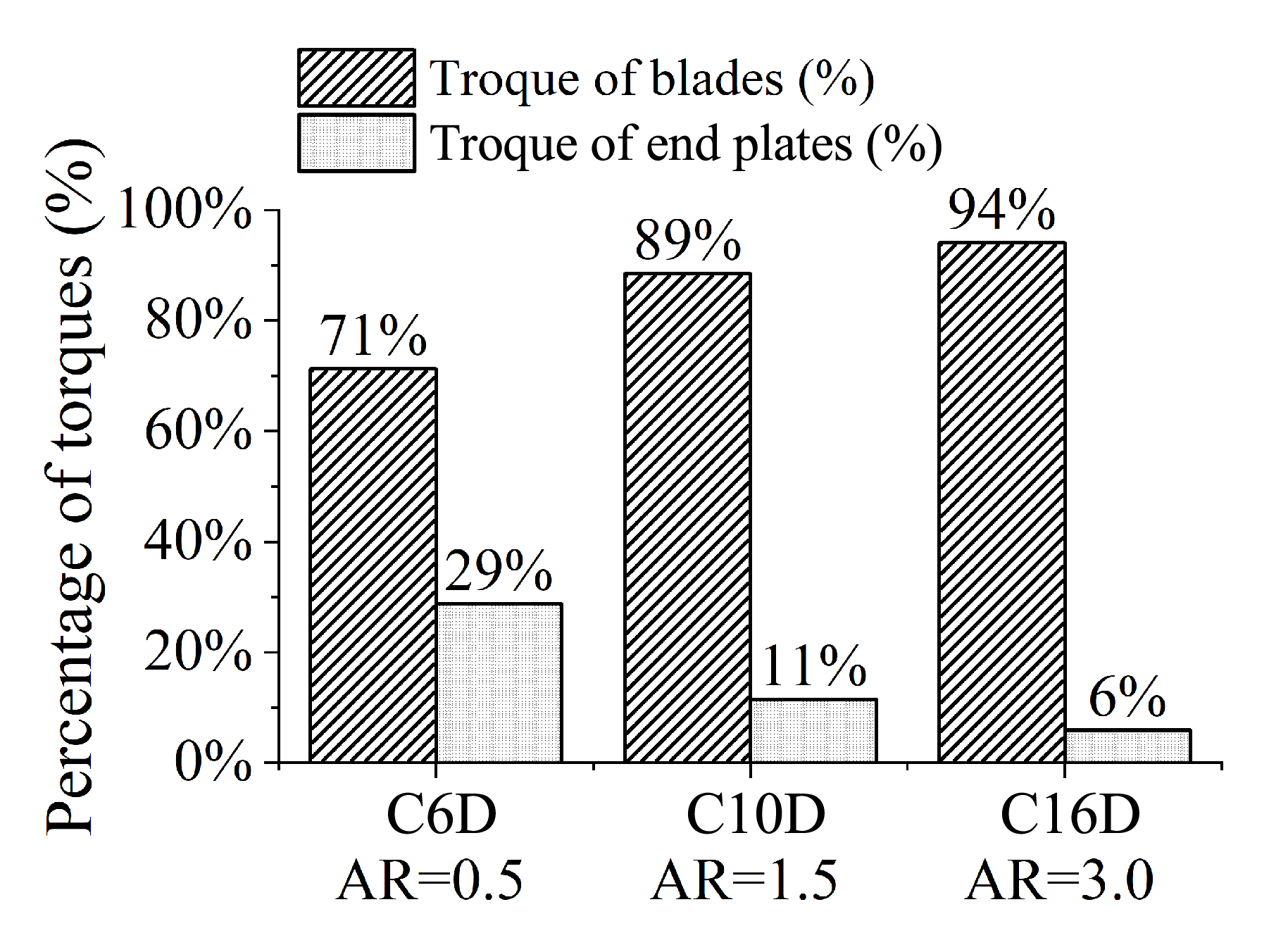}
        \label{fig:C6DC10DC16D_TorqueRatio}
    }
    \caption{Comparison of torques for cases C6D, C10D and C16D \\ (Obtained by numerical simulations, $\alpha_{max} = 40^\circ$)}
    \label{fig:Case_C6DC10DC16D_Q_All}
\end{figure}

Numerical simulations are performed based on cases C6D ($AR=0.5$), C10D ($AR=1.5$), and C16D ($AR=3.0$). Their dimensions and configurations are identical, except for the blade aspect ratio. The torque forces produced by the blades and end plates are extracted from the numerical simulation results, as shown in Figure.\ref{fig:Case_C6DC10DC16D_Q_All}. It can be seen that the friction force on the rotating end plates introduces extra torque and hence results in reduced aerodynamic efficiency. Figure.\ref{fig:C6DC10DC16D_Torque} indicates that although the aspect ratio of the blade varies from $0.5$ to $3.0$, the torque produced by the rotating end plates remains constant at the given rotation speed. As shown in Figure.\ref{fig:C6DC10DC16D_TorqueRatio}, the portion of torque from the end plates increases with a reduction in the blade aspect ratio due to a decrease in the blade area. This leads to a deterioration in aerodynamic efficiency. Therefore, it is beneficial to keep the end plates stationary. 

\begin{figure}
    \centering
    \includegraphics[width=0.9\linewidth]{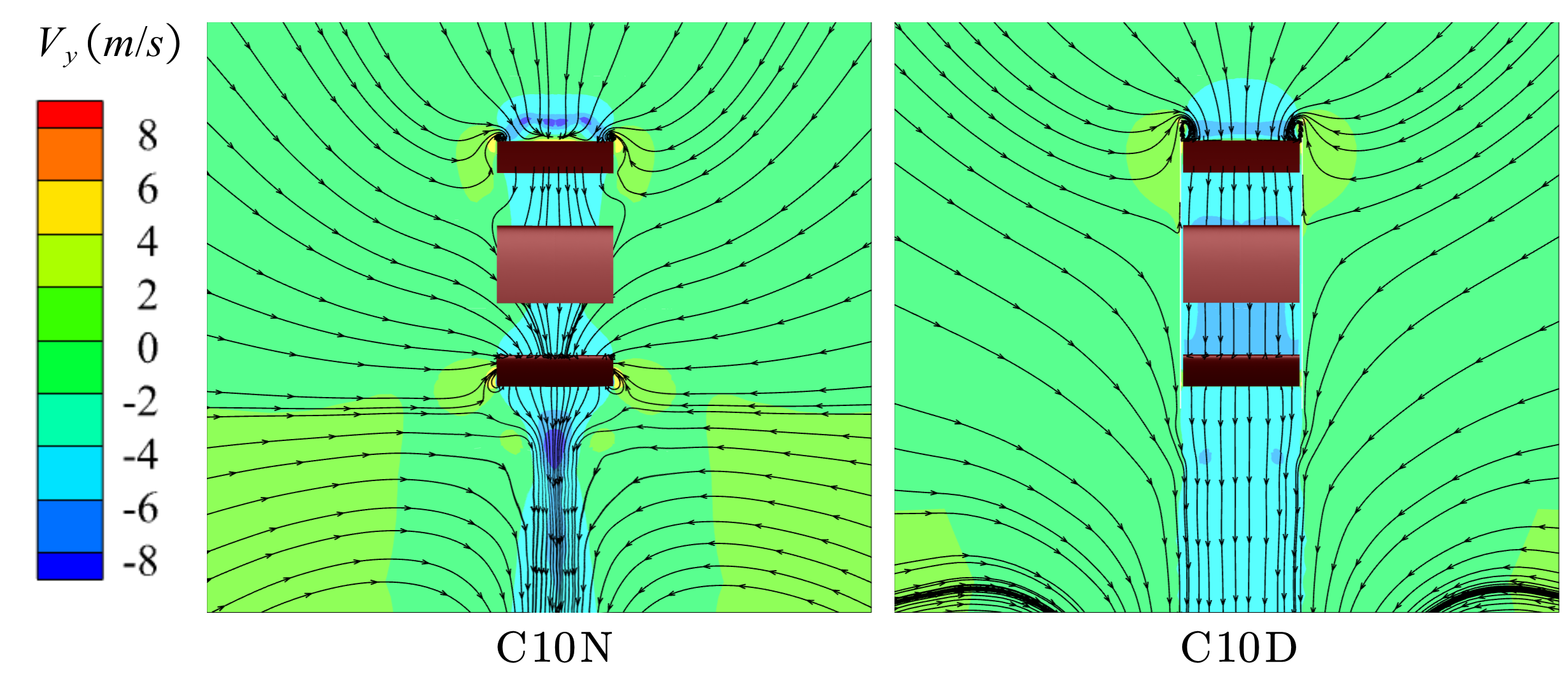}
    \caption{Slipstream of cycloidal propeller with and without end plates}
    \label{fig:contraction_RotorStream}
\end{figure}

The slipstreams of the rotors with and without end plates are shown in Figure.\ref{fig:contraction_RotorStream}.  For cases without end plates, strong contraction of the slipstream can be observed, which represents an induced power expenditure\cite{pereira2008hover}. With the help of end plates, the slipstream is forced to maintain a constant cross-sectional area; therefore, the aerodynamic efficiency can be improved.  

\subsection{Effects of blade aspect ratio for the cases with stationary end plates}

As depicted in the previous section, it is preferable to keep the end plates stationary to improve aerodynamic efficiency. Therefore, additional experiments are performed on the basis of stationary end plates. 

All cases defined in Table.\ref{table:Experiment_Cases_SEP}  are derived from the cases in Table.\ref{table:Experiment_Cases_AR}. These cases share the same dimensions as their counterparts, except they have fixed end plates. As an example, case SC6D serves as a modified version of case C6D, incorporating two static end plates.

\begin{table}[ht]\centering
\caption{Parameters of the cycloidal propellers with stationary thin end plates}
\begin{tabular}{ccccc}
\hline
Case ID & $AR$ & $t_e$ &$\delta_t$\\
\hline
SC6D  & $0.50$ & $2mm$ & $4mm$\\
SC8D  & $1.00$ & $2mm$ & $4mm$\\
SC10D & $1.50$ & $2mm$ & $4mm$\\
SC12D & $2.00$ & $2mm$ & $4mm$\\
SC14D & $2.50$ & $2mm$ & $4mm$\\
SC16D & $3.00$ & $2mm$ & $4mm$\\
\hline
\label{table:Experiment_Cases_SEP}
\end{tabular}
\end{table}

\begin{figure}
    \centering
    \includegraphics[width=0.9\linewidth]{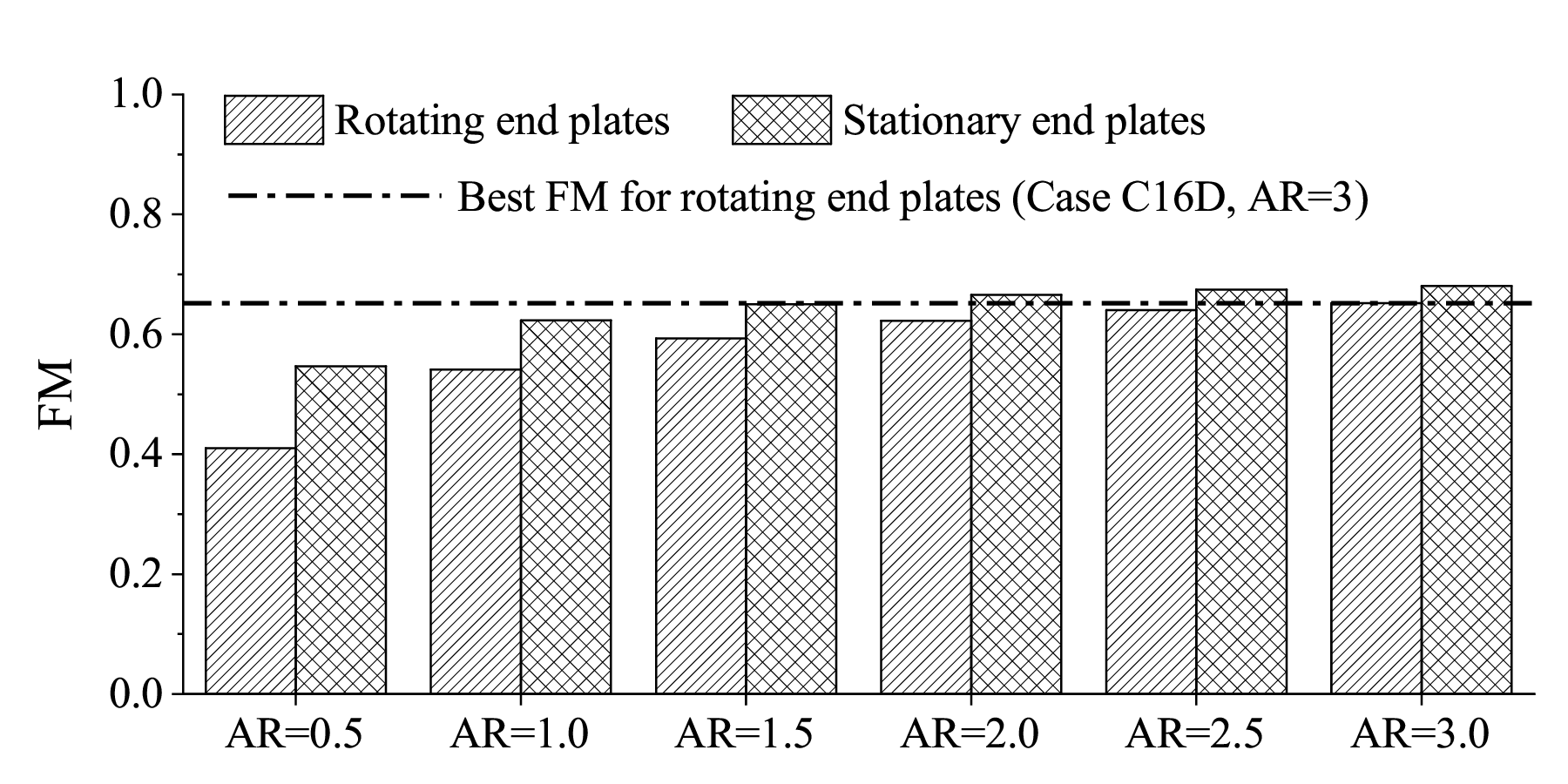}
    \caption{Effects of blade aspect ratio for the cases with rotating and stationary end plates}
    \label{fig:Static_Case_FM}
\end{figure}

The hovering efficiency of these cases is shown in Figure.\ref{fig:Static_Case_FM}. It can be seen that the cases with stationary end plates outperform their counterparts with rotating end plates. The hovering efficiency of the SC10D case ($AR=1.5$) already matches that of the C16D case, although its $AR$ is half that of the C16D case. 

\begin{figure}
    \centering
    \subfigure[$C_{T}$ vs. Azimuth angle]
    {
        \includegraphics[width=0.47\linewidth]{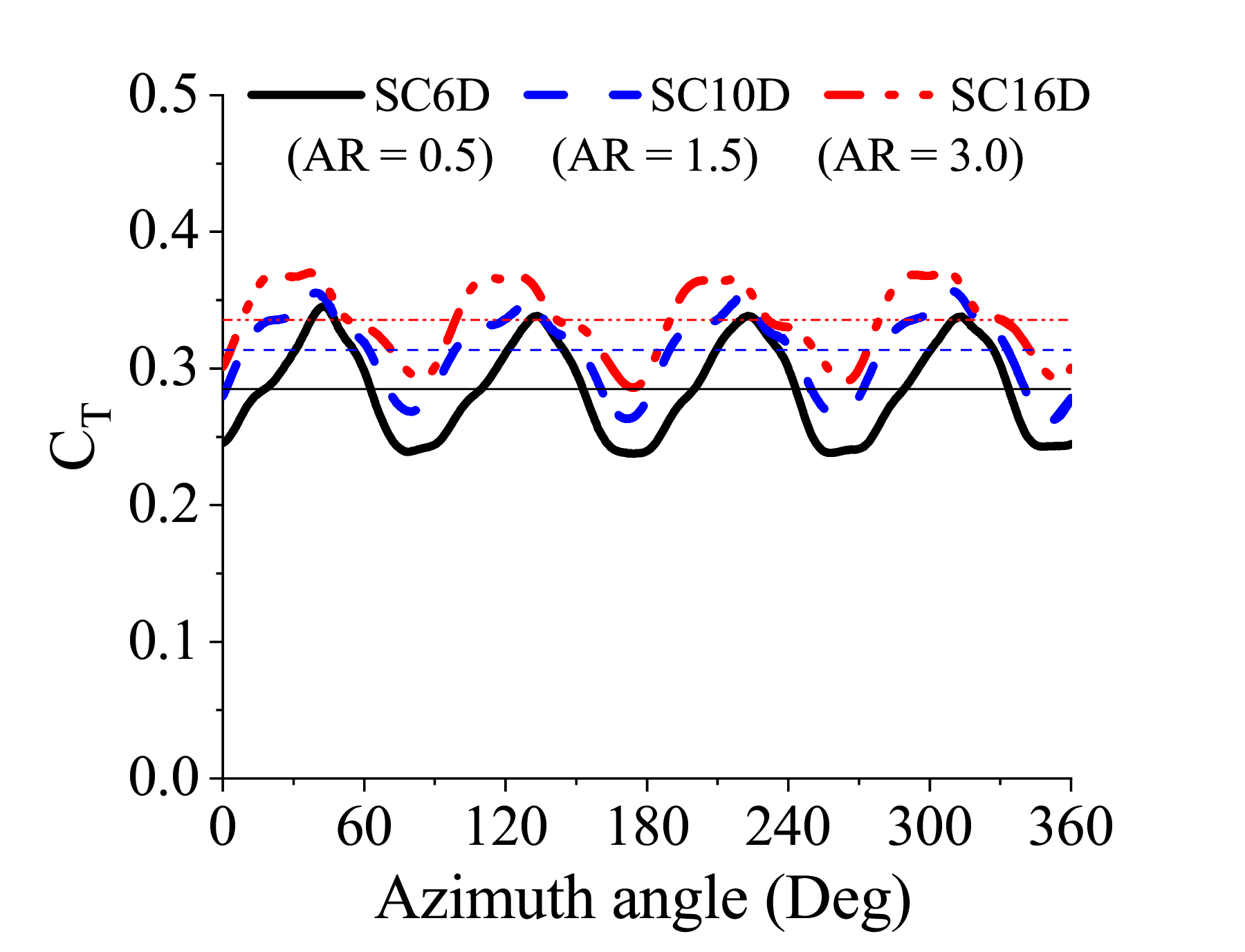}
        \label{fig:Angle_Fn_Total_AR}
    }
    \subfigure[$C_{Q}$ vs. Azimuth angle]
    {
        \includegraphics[width=0.47\linewidth]{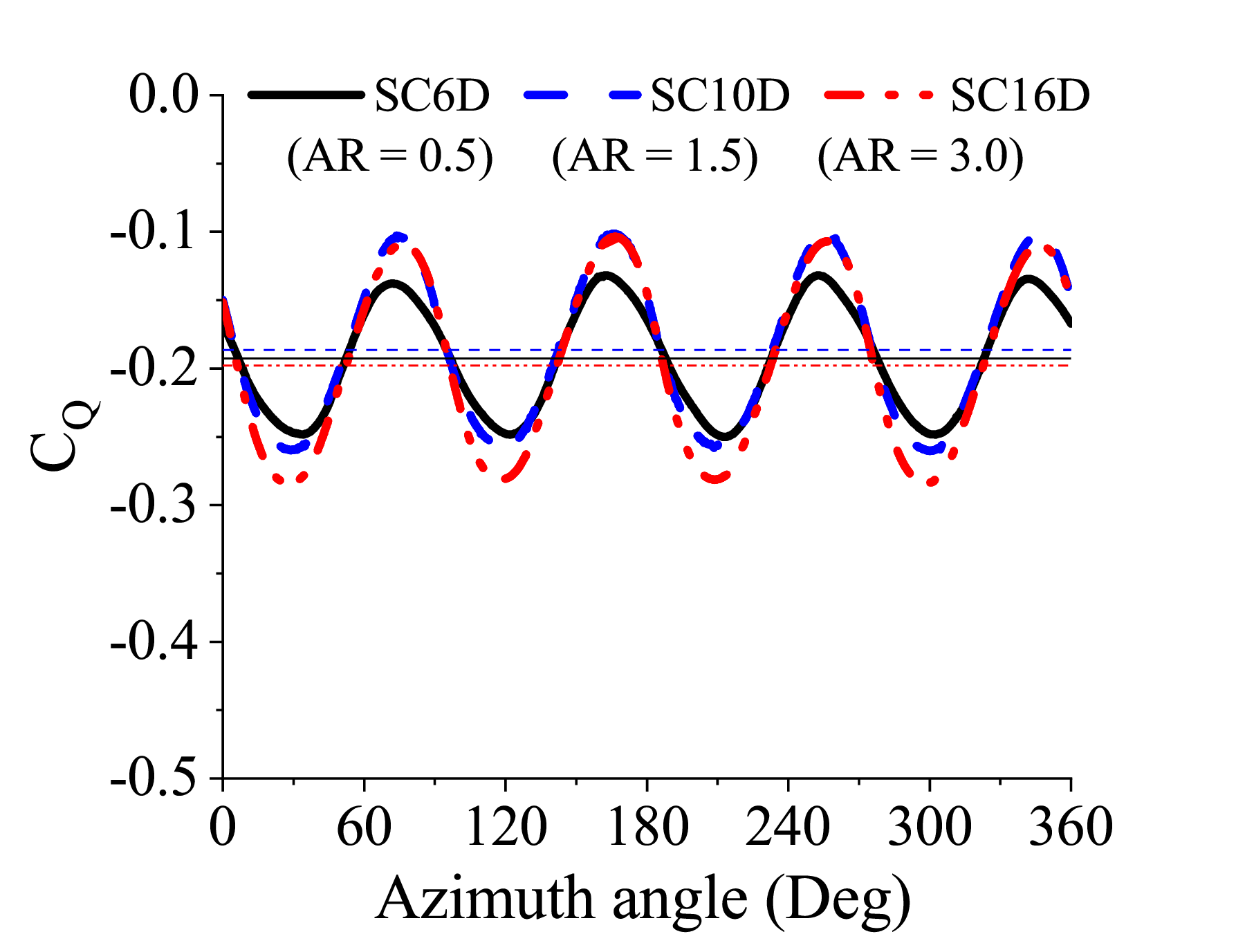}
        \label{fig:Angle_Torque_Total_AR}
    }
    \caption{The aerodynamic force of the SC6D, SC10D, and SC16D over one cycle (Obtained by numerical simulations)}
    \label{fig:StationaryDataCompare}
\end{figure}

However, the hovering efficiency still decreases with the blade aspect ratio, although the end plates are stationary and do not draw extra power. Figure. \ref{fig:StationaryDataCompare} compares the aerodynamic force data of the SC6D ($AR=0.5$), SC10D ($AR=1.5$), and SC16D ($AR=3.0$). It can be seen that these cases have similar magnitudes of $C_{Q}$ but show a reduction in $C_{T}$. This results in a reduction in hovering efficiency. 

\begin{figure}
    \centering
    \includegraphics[width=0.97\linewidth]{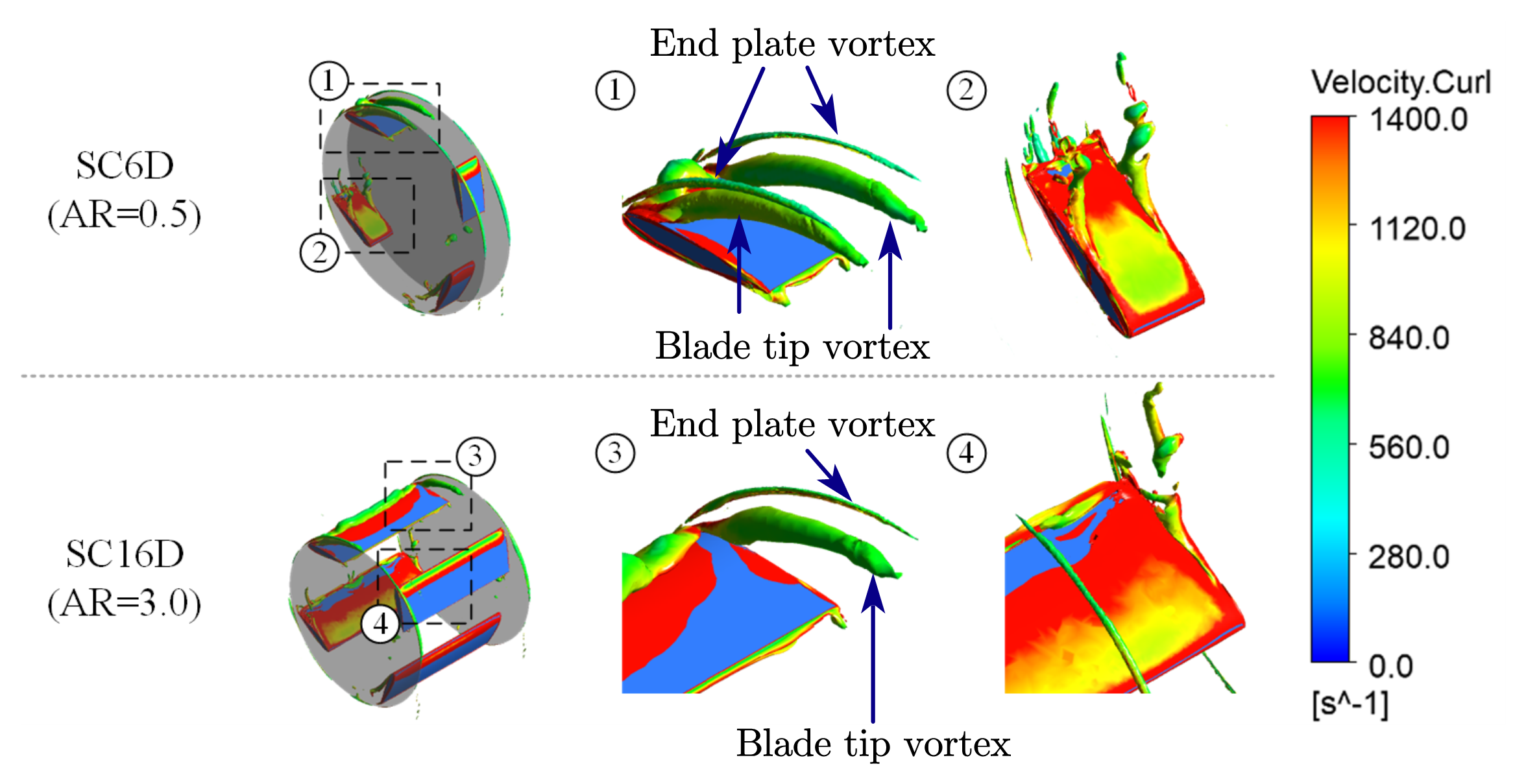}
    \caption{The end plate vortex and blade tip vortex of SC6D and SC16D ($\theta=216^{\circ}$)}
    \label{fig:Velocity_Curl}
\end{figure}

This is because there is a gap between the blade tip and end plate ($\delta_t=4mm$ in all cases). This gap leads to the formation of blade tip vortices (BTV) near the blade tips. The BTV produced by blades with a lower aspect ratio is stronger than that produced by blades with a higher $AR$, as shown in Figure.\ref{fig:Velocity_Curl}. The presence of BTV causes tip loss and reduces the effective blade span. For cases with a lower blade aspect ratio, a larger portion of the blade area is impacted by the BTV, which, in turn, decreases the overall thrust coefficient. Additionally, when the aspect ratio is low, the stronger BTV generates greater induced drag. On the other hand, the blades with smaller $AR$ have a smaller surface area and, hence, smaller friction drag. Consequently, the torque coefficient remains almost unchanged. Therefore, it is crucial to keep the blade tip gap as small as possible. 

\subsection{Effects of end plate thickness for the cases with stationary end plates}

\begin{figure}
    \centering
    \includegraphics[width=0.6\linewidth]{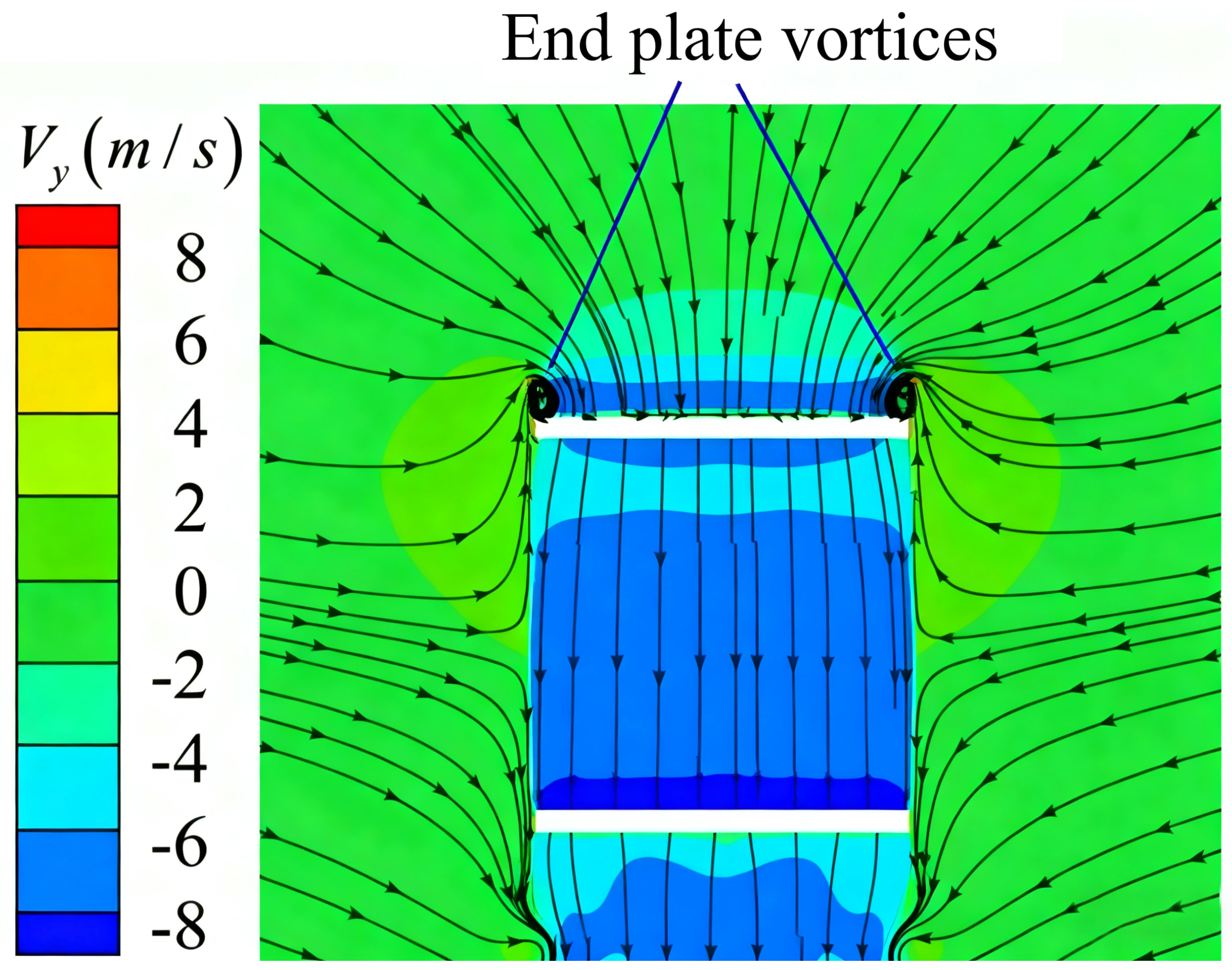}
    \caption{The end plate vortices generated by thin end plates}
    \label{fig:SD_End_plate_vortices}
\end{figure}

In Figure.\ref{fig:SD_End_plate_vortices}, streamlines are illustrated at the cross-section of a cycloidal propeller with a thin-walled end plate. As the airflow is drawn into the top of the propeller disk, an end plate vortex is generated inside the top of the end plate, leading to a reduction in blade thrust. To tackle the adverse effects of the end plate vortex on hovering efficiency, thick end plates with rounded edges are introduced. The corresponding experimental setups are detailed in Table.\ref{table:Experiment_Cases_SCDT}, with a constant pitch amplitude of $40\deg$ and an end plate thickness of $25mm$ across all cases. These cases possess identical geometrical configurations to those defined in Table \ref{table:Experiment_Cases_SEP}, with the exception of the end plate thickness.

\begin{table}[ht]\centering
\caption{Parameters of the cycloidal propellers with stationary thick end plates}
\begin{tabular}{ccccc}
\hline
Case ID & $AR$ &$t_e$ & $\delta_t$\\
\hline
SC6DT  & $0.50$ & $25mm$ & $4mm$ \\
SC8DT  & $1.00$ & $25mm$ & $4mm$ \\
SC10DT & $1.50$ & $25mm$ & $4mm$ \\
SC12DT & $2.00$ & $25mm$ & $4mm$ \\
SC14DT & $2.50$ & $25mm$ & $4mm$ \\
SC16DT & $3.00$ & $25mm$ & $4mm$ \\ 
\hline
\label{table:Experiment_Cases_SCDT}
\end{tabular}
\end{table}

\begin{figure}
    \centering
    \includegraphics[width=0.9\linewidth]{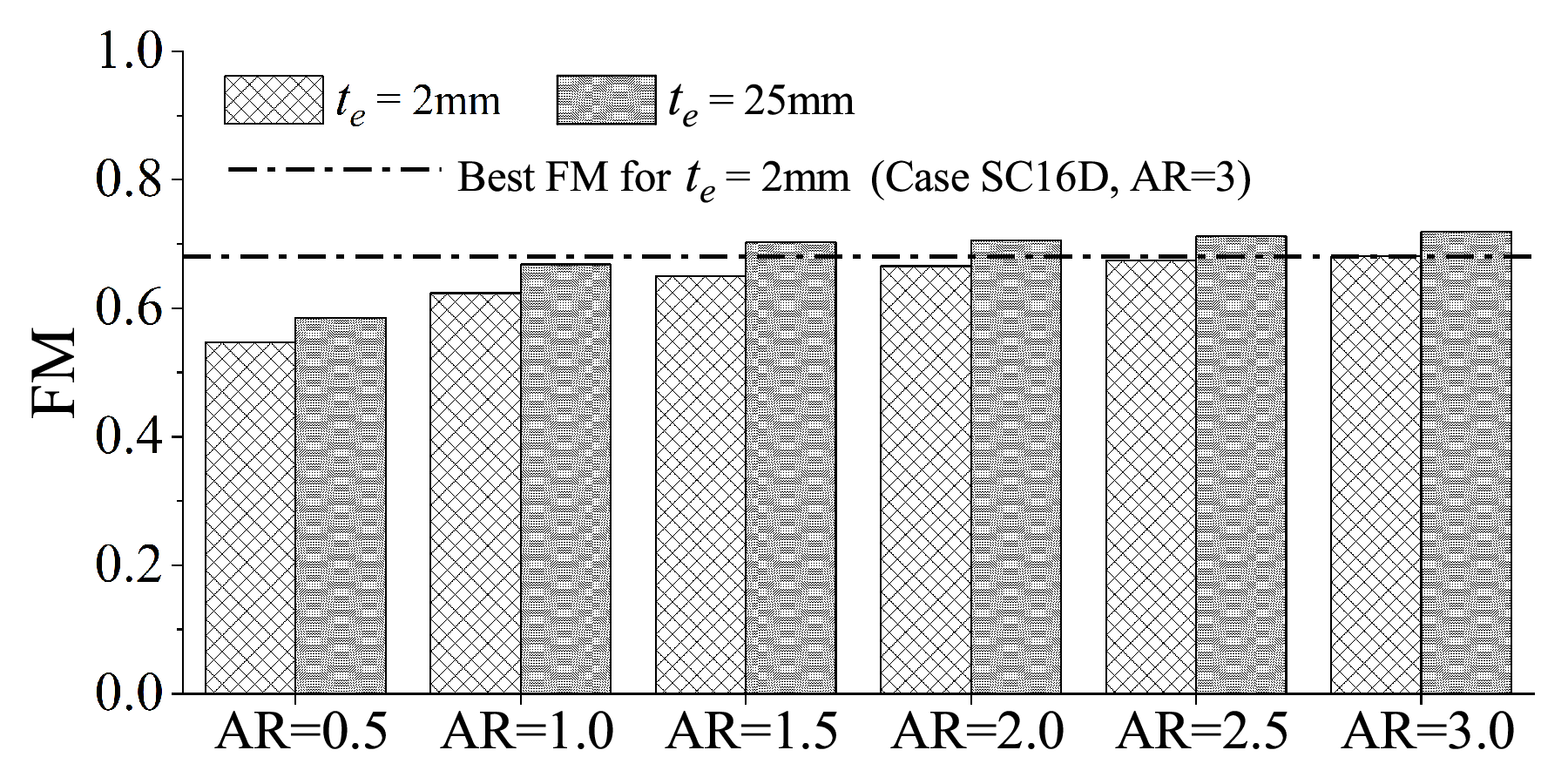}
    \caption{Comparison of $FM$ for the cases with thin and thick end plates}
    \label{fig:SCDT_EndPlates_Thinckness_FM}
\end{figure}

As depicted in Figure.\ref{fig:SCDT_EndPlates_Thinckness_FM}, the cycloidal propeller with thick fixed end plates shows superior hovering efficiency compared to those with thin end plates when their aspect ratios are equal. The maximum efficiency of the design with thick end plates is 0.72, which is comparable to that of helicopter rotors. For the cases with $AR\geq1.5$, the hovering efficiency remains almost unchanged, regardless of the end plate thickness. This suggests that the design with $AR = 1.5$ can be adopted to reduce the weight of the blade structure.

\begin{figure}
    \centering
    \subfigure[Definition of the slices]
    {
        \includegraphics[width=0.45\linewidth]{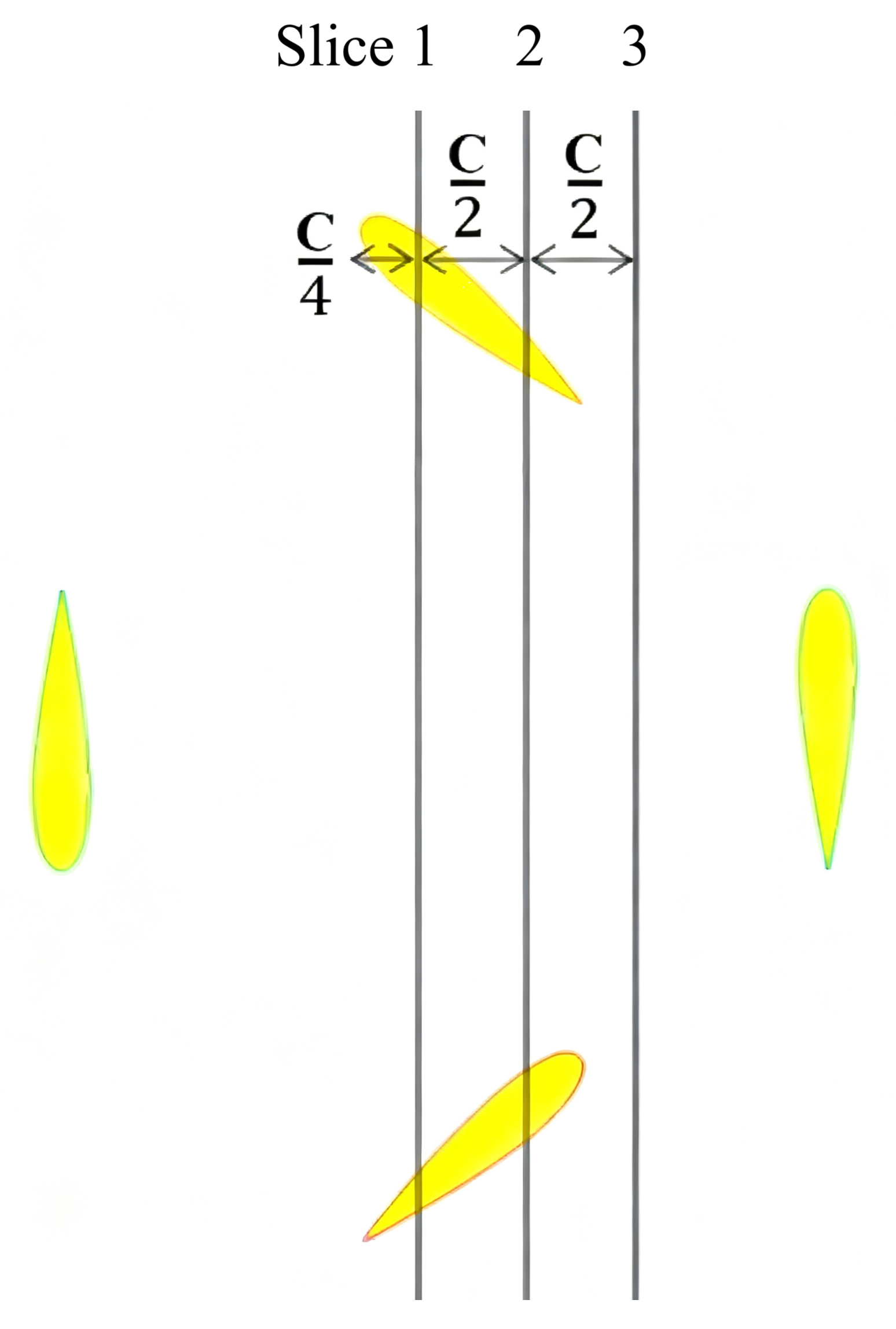}
        \label{fig:SCD_SCDT_Slices_a}
    }
    \subfigure[Streamlines on each slice]
    {
        \includegraphics[width=0.48\linewidth]{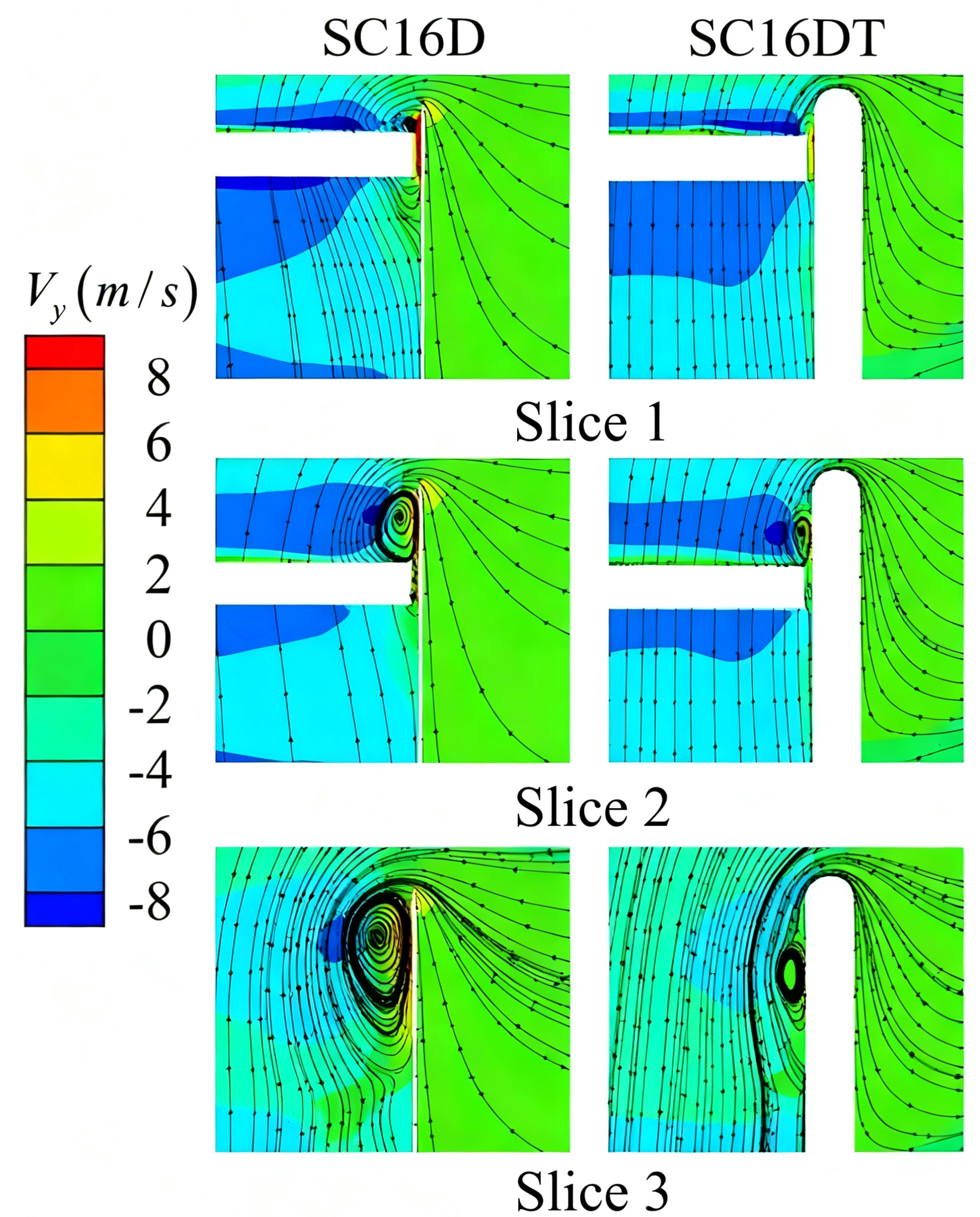}
        \label{fig:SCD_SCDT_Slices_b}
    }
    \caption{Comparison of the streamlines along the upper edges of end plates with different thickness (Obtained by numerical simulations)}
    \label{fig:ThicknesStreamlineCompare}
\end{figure}

\begin{figure}
    \centering
    \subfigure[$C_T$ of the cycloidal propeller in one cycle]
    {
        \includegraphics[width=0.47\linewidth]{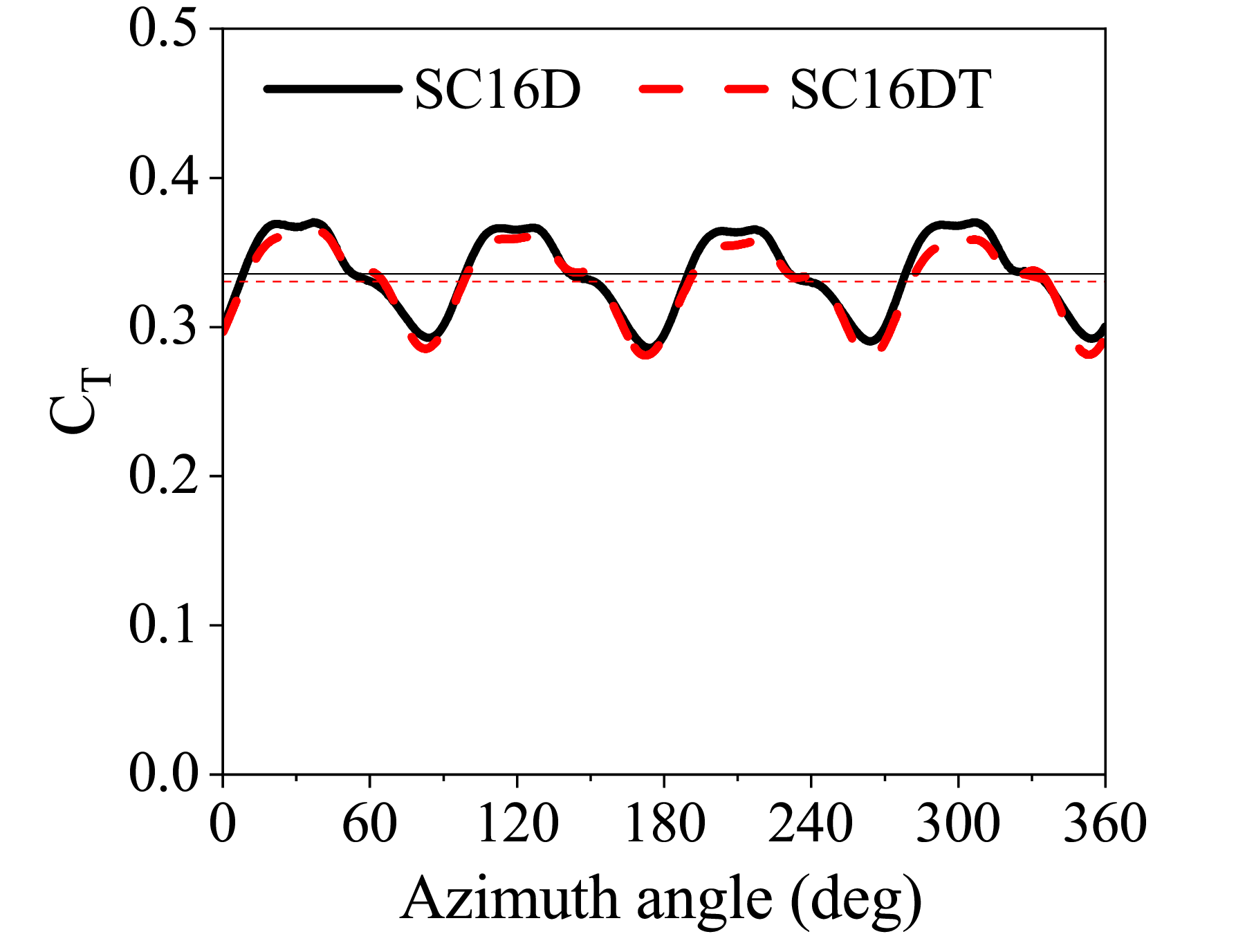}
        \label{fig:SCD_SCDT_CT}
    }
    \subfigure[$C_Q$ of the cycloidal propeller in one cycle]
    {
        \includegraphics[width=0.47\linewidth]{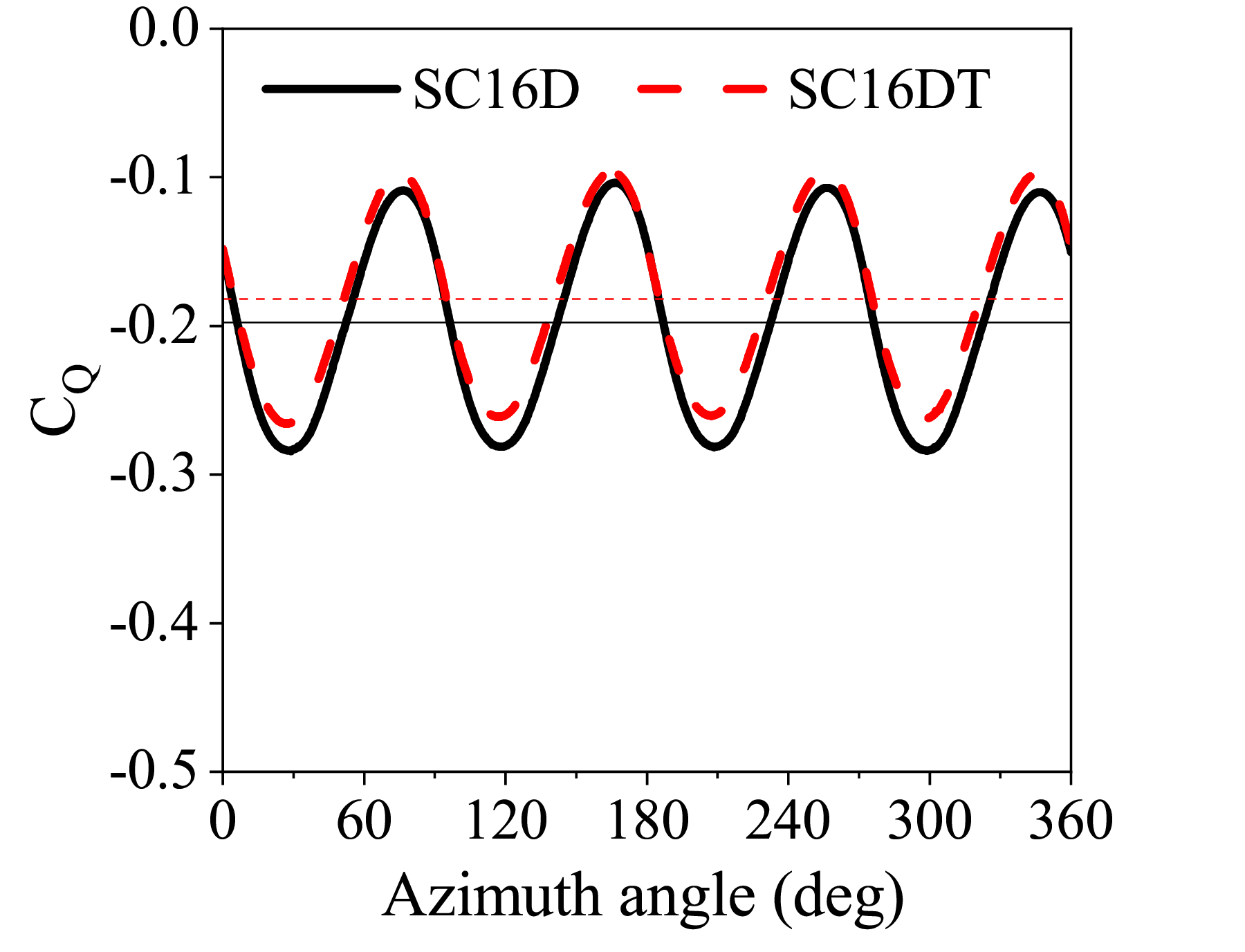}
        \label{fig:SCD_SCDT_CQ}
    }
    \caption{Comparison of aerodynamic forces for the end plates with different thickness (Obtained by numerical simulations)}
    \label{fig:ThicknesCTCQCompare}
\end{figure}

Numerical simulations were conducted based on the cases SC16D and SC16DT. Figure \ref{fig:ThicknesStreamlineCompare} illustrates the streamlines across various slices. The case with thick end plates (SC16DT) features a larger end plate edge radius. Hence, the intensity of the end plate vortices is significantly reduced. This decreases the aerodynamic torque, thereby improving the hovering efficiency, as shown in Figure. \ref{fig:ThicknesCTCQCompare}.

Whereas for the case featuring a thin end plate (SC16D ), an end plate vortex arises at the blade's leading edge. The end plate vortex merges with the tip vortex at the blade trailing edge, creating a composite vortex structure that interacts with downstream blades, producing stronger downwash in the rotor cage and larger torque, resulting in lower hovering efficiency. 

\section{Conclusions}

In this paper, cycloidal propellers with end plates are studied based on force measurement experiments, and the details of aerodynamic forces and the flow field are analyzed based on numerical simulations. The following conclusions can be reached:

\begin{enumerate}[label=(\arabic*). ]

\item Experiment results indicate that the cycloidal propellers with two stationary thick end plates, chord-to-radius ratio of $0.65$, and a large pitching amplitude of $40^{\circ}$ outperform other designs. It reaches a hovering efficiency of $0.72$ with a blade aspect ratio of $3$. In contrast, the highest $FM$ achieved by the cases without end plates is merely $0.54$. 
\par
\item Because the end plates can suppress the tip vortex and reduce the induced loss, they result in a significantly better hovering efficiency over the designs without end plates.
\par
\item Utilizing stationary thick end plates helps maintain hovering efficiency even when the blade aspect ratio is very small. As a result, hovering efficiency can exceed $0.7$ despite the blade aspect ratio being merely $1.5$. Decreasing the blade aspect ratio can address the structural issues due to centrifugal forces, and hence can reduce the structure weight.
\par
\item Cycloidal propellers with stationary end plates are superior to the designs with rotating end plates, since rotating end plates will introduce extra torque due to aerodynamic friction force.
\par
\item Designs featuring thick end plates outperform their counterparts with thin end plates since the round edges contribute to reducing end plate vortices.
\par
\item $FM$ decreases with a lower blade aspect ratio in both rotating and stationary end plate cases. However, the decrease in $FM$ is less significant when using stationary end plates.

\end{enumerate}

\section{Acknowledgment}
The work was supported by the National Natural Science Foundation of China (Grant No. 52272380).


\nocite{*}
\bibliography{References}

\end{document}